\documentclass[lettersize,journal]{IEEEtran}
\usepackage{amsmath,amsfonts,amsthm,amssymb}
\usepackage{algorithm,algorithmic}
\usepackage{array}
\usepackage{booktabs}    
\usepackage[table]{xcolor}
\usepackage{xcolor}
\usepackage{bbm}

\usepackage{textcomp}
\usepackage{color}
\usepackage{stfloats}

\usepackage{url}
\usepackage{verbatim}
\usepackage{cite}

\usepackage{float} 
\usepackage{graphicx}

\usepackage{caption}
\usepackage[caption=false,font=footnotesize,labelfont=rm,textfont=rm]{subfig}
\DeclareSubrefFormat{myparens}{#1(#2)}
\newcolumntype{L}{>{\hspace*{-\tabcolsep}}l}
\newcolumntype{R}{c<{\hspace*{-\tabcolsep}}}
\definecolor{lightblue}{rgb}{0.93,0.95,1.0}

\def\BibTeX{{\rm B\kern-.05em{\sc i\kern-.025em b}\kern-.08em
	T\kern-.1667em\lower.7ex\hbox{E}\kern-.125emX}}
\usepackage{balance}

\makeatletter

\newcommand{\Rmnum}[1]{\expandafter\@slowromancap\romannumeral #1@}
\makeatother

\theoremstyle{remark}

\newcommand{\cA}{\mathcal{A}}

\newcommand{\cK}{\mathcal{K}}

\newcommand{\cM}{\mathcal{M}}
\newcommand{\cN}{\mathcal{N}}
\newcommand{\cO}{\mathcal{O}}
\newcommand{\cP}{\mathcal{P}}

\newcommand{\cR}{\mathcal{R}}

\newcommand{\cU}{\mathcal{U}}

\newcommand{\cX}{\mathcal{X}}

\newcommand{\ba}{\mathbf{a}}

\newcommand{\be}{\mathbf{e}}
\newcommand{\bbf}{\mathbf{f}}

\newcommand{\bj}{\mathbf{j}}

\newcommand{\bbm}{\mathbf{m}}
\newcommand{\bn}{\mathbf{n}}

\newcommand{\bp}{\mathbf{p}}

\newcommand{\br}{\mathbf{r}}
\newcommand{\bs}{\mathbf{s}}

\newcommand{\bx}{\mathbf{x}}
\newcommand{\by}{\mathbf{y}}
\newcommand{\bz}{\mathbf{z}}

\newcommand{\bA}{\mathbf{A}}
\newcommand{\bB}{\mathbf{B}}

\newcommand{\bF}{\mathbf{F}}
\newcommand{\bG}{\mathbf{G}}
\newcommand{\bH}{\mathbf{H}}
\newcommand{\bI}{\mathbf{I}}
\newcommand{\bJ}{\mathbf{J}}

\newcommand{\bR}{\mathbf{R}}

\newcommand{\bT}{\mathbf{T}}
\newcommand{\bU}{\mathbf{U}}

\newcommand{\bX}{\mathbf{X}}

\newcommand{\remi}{\mathrm{emi}}
\newcommand{\rhw}{\mathrm{hw}}
\newcommand{\rd}{\mathrm{d}}

\newcommand{\bbC}{\mathbb{C}}
\newcommand{\bbR}{\mathbb{R}}

\newcommand{\bbZ}{\mathbb{Z}}

\newcommand{\bzero}{\mathbf{0}}

\newcommand{\bSigma}{{\boldsymbol\Sigma}}

\newcommand{\bLambda}{{\boldsymbol\Lambda}}

\newcommand{\bxi}{{\boldsymbol\xi}}
\newcommand{\bXi}{{\boldsymbol\Xi}}

\newcommand{\bkappa}{\boldsymbol{\kappa}}

\newcommand{\bbone}{\mathbbm{1}}
\newcommand{\dr}{\mathrm{d}\mathbf{r}}
\newcommand{\ds}{\mathrm{d}\mathbf{s}}

\newcommand{\figref}[1]{Fig.~\ref{#1}}
\newcommand{\alref}[1]{\textbf{Algorithm}~\textbf{\ref{#1}}}

\newcommand{\rsum}{\mathrm{sum}}
\newcommand{\st}{\mathrm{s.t.}}

\newcommand{\rdiag}[1]{\mathrm{diag}\left\{ #1\right\}}
\newcommand{\trace}[1]{\mathrm{tr}\left(#1\right)}
\newcommand{\expect}[1]{\mathbb{E}{\left\{#1\right\}}}

\newcommand{\real}[1]{\Re\{#1\}}

\begin{document}
\title{On the Spectral Efficiency of Multi-user Holographic MIMO Uplink Transmission}

\author{
	Mengyu~Qian, \IEEEmembership{Graduate Student Member,~IEEE,} 
	Li~You,~\IEEEmembership{Senior Member,~IEEE,} 
	Xiang-Gen~Xia,~\IEEEmembership{Fellow,~IEEE,}
	and~Xiqi~Gao,~\IEEEmembership{Fellow,~IEEE}
	\thanks{Part of this work was submitted to 2024 IEEE Global Communications Conference (GLOBECOM) \cite{qianmengyu_HMIMO_conference}.}%
	\thanks{Mengyu Qian, Li You, and Xiqi Gao are with the National Mobile Communications Research Laboratory, Southeast University, Nanjing 210096, China, and also with the Purple Mountain Laboratories, Nanjing 211100, China (e-mail: qianmy@seu.edu.cn; lyou@seu.edu.cn; xqgao@seu.edu.cn).}
	\thanks{Xiang-Gen Xia is with the Department of Electrical and Computer Engineering, University of Delaware, Newark, DE 19716 USA (e-mail: xianggen@udel.edu).}
}

\maketitle

\begin{abstract}
With antenna spacing much less than half a wavelength in confined space, holographic multiple-input multiple-output (HMIMO) technology presents a promising frontier in next-generation mobile communication. We delve into the research of the multi-user uplink transmission with both the base station and the users equipped with holographic planar arrays. To begin, we construct an HMIMO channel model utilizing electromagnetic field equations, accompanied by a colored noise model that accounts for both electromagnetic interference and hardware noise. Since this model is continuous, we approximate it within a finite-dimensional space spanned by Fourier space series, which can be defined as the communication mode functions. We show that this channel model samples Green's function in the wavenumber domain in different communication modes. Subsequently, we tackle the challenging task of maximizing the spectral efficiency (SE) of the system, which involves optimizing the continuous current density function (CDF) for each user. Using the aforementioned approximation model, we transform the optimization variables into expansion coefficients of the CDFs on a finite-dimensional space, for which we propose an iterative water-filling algorithm. Simulation results illustrate the efficacy of the proposed algorithm in enhancing the system SE and show the influence of the colored noise and the system parameters on the SE.
\end{abstract}

\begin{IEEEkeywords}
	Holographic MIMO communications, electromagnetic modeling,  Fourier space series,  colored noise, spectral efficiency.
\end{IEEEkeywords}

\section{Introduction}\label{sec:introduction}
As the realm of mobile communication extends into the fifth generation (5G) and beyond, researchers are fervently exploring next-generation solutions, as highlighted in recent research \cite{youTowards6G,wang2024electromagnetic}. The anticipated sixth generation (6G) is envisioned as a colossal distributed neural network, seamlessly integrating communication, perception, and computation capabilities, heralding the era of the ``Internet of Everything'' \cite{10158439, you6GTextTKMu2023,ICL_Qiang,jin2024i2i,zhu2024ubi}. To fulfill these visionary 6G communication goals, researchers and engineers face the unrelenting challenge of meeting ever-increasing demands for higher data rates, broader coverage, and enhanced quality of service \cite{10054381, dangWhatShould6G2020,jiang2023rate}.

In response to these burgeoning requirements, the exploration of innovative communication technologies has become imperative \cite{9847609,yeFluidAntennaassistedMIMO2023,xuNearFieldWidebandExtremely2022a,huangkelin,jiang2023total,jiang2022hybrid,hu2018beyond}. One of the most promising avenues is holographic multiple-input multiple-output (HMIMO), also known as continuous-aperture MIMO (CAP-MIMO) \cite{2022arXiv221201257G,huangHolographicMIMOSurfaces2020,damicoHolographicMIMOCommunications2023,zhao2024performance,castellanos2023electromagnetic,gongHolographicMIMOCommunications2024a}. This approach harnesses the unique capabilities of MIMO antenna systems in a novel and sophisticated manner \cite{pizzoHolographicMIMOCommunications2020, wanTerahertzMassiveMIMO2021,xuNearFieldWidebandExtremely2022a,gongNearFieldChannelModeling2024a}. Unlike traditional MIMO or massive MIMO, the spacing between each antenna in a holographic array is significantly smaller than half a wavelength, allowing for a denser and even quasi-continuous deployment of antennas within limited space \cite{an2023tutorial, zhu2024electromagnetic}. Under such conditions, proper modeling of electromagnetic (EM) propagation becomes a fundamental consideration \cite{pizzoSpatiallyStationaryModelHolographic2020}. Conventional channel models may not accurately represent the wireless link characteristics concerning path loss and the number of  orthogonal EM channels (communication modes) when using holographic arrays \cite{dardariHolographicCommunicationUsing2021}.

Furthermore, classical models rely on distinct current distributions without accounting for the holographic capability offered by metasurfaces \cite{wei2022multi2}. Therefore, models adopted by HMIMO communication must incorporate this design flexibility, relying on the transmission of information through EM waves within the continuous wireless channels (EM information theory) \cite{anTutorialHolographicMIMO2023a, anTutorialHolographicMIMO2023b, zhangCapacityAnalysisHolographic2023}.

Optimal transmission between holographic arrays considers communication between continuous regions (volumes), including
extremely-dense antennas and the continuous wireless channel, as discussed in \cite{wei2022multi}.
This is different from the conventional MIMO model featuring discrete antennas, which samples the continuous-space EM channel according to the specific arrangements of the array elements \cite{wang2023tutorial}. 
Holographic communication becomes a challenge in functional analysis that relies primarily on geometric associations. The goal is to identify the most efficient sets of EM functions at both the transmitting and receiving ends to facilitate information transmission across various volumes. This approach allows exploration of the fundamental boundaries of communication, such as the inherent capacity of the continuous-space wireless channel, regardless of the specific technology and quantity of antenna elements \cite{dardariHolographicCommunicationUsing2021}.

To date, numerous researchers have conducted studies on communication with holographic arrays. A starting point for investigating denser arrays is to determine if they can provide larger degrees of freedom (DoFs), a topic explored by many scholars \cite{bucciDegreesFreedomScattered1989, bucciRepresentationElectromagneticFields1998, 9798854, brady2013beamspace}. For instance, in \cite{dardariCommunicatingLargeIntelligent2020}, the authors derived accurate analytical formulations for link gain and spatial DoFs from the EM perspective. In \cite{dingDegreesFreedom3D2022}, researchers defined a receiving coordinate system and explored the effect of the 3D array position  and rotation on achievable spatial DoFs in the linear propagation channel. Considering the  area and geometry constraints of the antenna array,  \cite{poonDegreesFreedomMultipleantenna2005} investigated the limitations on the number of spatial DoFs available to multi-antenna systems, including cases with both polarized and multi-polarized antennas.

Additionally, establishing channel model that accurately reflects physical reality and exploring their new features are areas ripe for in-depth research \cite{pizzoSpatialCharacterizationElectromagnetic2022, pizzoFourierPlaneWaveSeries2022, chen2024near,li2023electromagnetic}. In \cite{wei2022multi}, researchers demonstrated that EM channel models for multi-user HMIMO systems should not rely on i.i.d. Rayleigh fading modeling. Furthermore, there is keen interest in understanding how the increasing continuity of arrays can benefit the system. For instance, in \cite{jensenCapacityContinuousSpaceElectromagnetic2008}, the authors analyzed the capacity bounds of the system by identifying optimal antenna characteristics that maximize capacity for specific propagation scenarios. In \cite{wanMutualInformationElectromagnetic2023}, researchers presented a scheme for analyzing performance limits between continuous transceivers. Additionally, in \cite{miglioreHorseElectromagneticsMore2019}, 
the authors adhered to Kolmogorov's theory to quantify the information capacity achievable by a wireless communication system, taking into account the physical constraints imposed by electromagnetism. In \cite{zhangCapacityAnalysisHolographic2023}, researchers analyzed HMIMO channel capacity under realistic angular distribution and array aperture constraints, computing the spectral density of the generalized angular distribution. In \cite{mikkiShannonInformationCapacity2023}, the authors explored the information capacity limit of generic EM surfaces characterized by Shannon's information theory. Moreover, in \cite{zhangPatternDivisionMultiplexingMultiUser2023}, researchers introduced a scheme called pattern division multiplexing to design patterns for CAP-MIMO systems to maximize the total system rate, while in \cite{sanguinettiWavenumberDivisionMultiplexingLineofSight2023}, researchers introduced a wavenumber-domain multiplexing approach primarily focused on one-dimensional transmitting and receiving regions, which, while not optimal, can be efficiently implemented.

\begin{table}[!t]
\renewcommand\arraystretch{1.2}
\footnotesize
\caption{Notation List}
\label{table1}
\centering
\setlength{\tabcolsep}{0.05mm}{
	\begin{tabular*}{1.005\linewidth}{LccR}
			\toprule
			Notation && Definition\\
			\midrule 
			\rowcolor{lightblue}
			$\triangleq$ && Definition\\
			$\bA$ && Matrix\\
			\rowcolor{lightblue}
			$\bx$ && Column vector\\
			$\bI_N$ &&$N\times N$ identity matrix\\
			\rowcolor{lightblue}
			$[\bA]_{m,n}$ &&The $(m,n)$th entry of $\bA$\\
			$\bzero$ && Zero matrix/vector\\
			\rowcolor{lightblue}
			$(\cdot)^H$ &&Conjugate transpose\\
			$\cA$ && Surface\\
			\rowcolor{lightblue}
			$|\cA|$ && Lebesgue measure of $\cA$\\
			$\mathcal{CN}(\ba,\bB)$ &\quad& Circular symmetric complex Gaussian distribution\\
			\rowcolor{lightblue}
			$\expect{}$ && Expectation\\
			$\trace{}$ && Matrix trace\\
			\rowcolor{lightblue}
			$\operatorname{diag}(\mathbf{a})$ && Diagonalization operator\\
			$\jmath = \sqrt{-1}$ && Imaginary unit\\
			\rowcolor{lightblue}
			$\det(\bA)$ && Determinant of $\bA$\\
			$()^{-1}$ && Matrix inverse\\
			\rowcolor{lightblue}
			$||\bx||$ && $\ell_2$ norm of a vector $\bx$\\
			$\bbZ^n$ && $n$-dimensional integer set\\
			\rowcolor{lightblue}
			$\bbR^n$ &&  $n$-dimensional real-valued number space\\
			$\cO$ && In the order of computation complexity\\
			\rowcolor{lightblue}
			$\hat{\be}_x$, $\hat{\be}_y$,  $\hat{\be}_z$ && Orthogonal vectors in Cartesian coordinates\\
			$\nabla \times \bF(\bx)$ &&Curl operation of a vector function $\bF(\bx)$\\
			\rowcolor{lightblue}
			$\real{}$ &&Real part\\
			$\mathbf{r}=r_x \hat{\be}_x+r_y\hat{\be}_y+r_z \hat{\be}_z$ && A point in $\bbR^3$\\
			\rowcolor{lightblue}
			$\lceil x \rceil$ && Taking the smallest integer greater than $x$\\
			$\bbone_\cX(x)$ && The indicator function over the set $\cX$\\
			\rowcolor{lightblue}
			$(a)^+$ && $\max\{a,0\}$\\
			$\odot$ && Hadamard product\\
			\bottomrule
		\end{tabular*}
	}
\end{table}

\textit{Motivation and Contributions:}
Most of the current research predominantly focuses on analyzing holographic arrays as one-dimensional line segments \cite{zhangPatternDivisionMultiplexingMultiUser2023,sanguinettiWavenumberDivisionMultiplexingLineofSight2023}. In other words, the arrays used by either the users or the base station, or sometimes both, are considered one-dimensional. Also, in the study of transmission systems assisted by holographic arrays, hardware noise and EM interference  are seldom addressed in combination. While it is convenient to describe the user end by approximating it as a point in three-dimensional space, such an approximation is no longer sufficient as the size of the user array increases and as communication moves towards far-near mixed fields and even near fields.
Furthermore, when analyzing system performance such as the system spectral efficiency (SE), most of the work assumes that the source current density is known, which is in general suboptimal. 
Therefore, there is a need for a comprehensive exploration of the channel characteristics of holographic planar arrays (HPAs)-assisted systems, as well as the optimization of the system SE.
To tackle these pertinent questions, it becomes imperative to employ EM field theory in establishing both the channel model and the colored noise model for the multi-user HPAs-assisted transmission system. This, in turn, allows us to formulate the problem of maximizing the  system SE.
Taking into account the  structure of this problem, we devise corresponding algorithms to address it effectively. Finally, through the use of simulations, we validate the impact of the colored noise on  the system SE and investigate how  various system parameters such as array size, propagation distance and wave frequency influence the system SE. 

Following this line, our contributions can be summarized as follows:

$\bullet$ Considering a multi-user uplink transmission system where the positions of users are arbitrary, we use unitary transformation to align each user's HPA to a plane parallel to the receiving HPA. Then, utilizing EM field equations, we develop both channel model and noise models for the uplink HPAs-assisted transmission system. This noise model not only incorporates hardware noise but also accounts for EM interference. Acknowledging the excessive computational complexity associated with the aforementioned models, we employ a Fourier space series expansion to approximate them, thereby enabling the mapping of high-dimensional physical quantities to a finite-dimensional space.

$\bullet$ We analyze the fundamental characteristics of an uplink transmission system aided by HPAs. We demonstrate that the degrees of freedom in HPAs-assisted systems exhibit an inverse relationship with propagation distance while maintaining a direct correlation with the size of the transmitting array. Besides, the finite-dimensional approximation of the channel model takes the form of a sampled representation of the dyadic Green's function. In the cases where the receiving array size approaches infinity, the receiving mode is completely determined by the transmitting mode.

$\bullet$ We formulate the problem of maximizing the spectral efficiency (SE) in a multi-user HPAs-assisted uplink transmission system. Leveraging the aforementioned finite-dimensional approximation, we transform the optimization variables into finite-dimensional matrices in discrete space, significantly simplifying the problem analysis. We propose an iterative water-filling algorithm to derive the final solution and validate its effectiveness through simulations.

\textit{Paper Outline and Notations:}
The structure of this paper is as follows.
In Section \ref{sec:electromagnetic MIMO model description}, we initiate our exploration by deriving the channel and noise models from the EM field equations.
Section \ref{sec:Fourier-PWF} delves into the approximation of continuous models by means of Fourier space series expansion, yielding a more manageable model. We further investigate the characteristics of the approximated channel model and highlight the distinctions between it and the continuous channel model.
Moving on to Section \ref{sec: Current density function design}, we employ the established models to formulate the problem of maximizing the system  SE and propose an iterative water-filling optimization algorithm to address this problem.
Section \ref{sec:Simulation} is dedicated to presenting simulation results that elucidate the relationship between the system SE and colored noise as well as system parameters.
For clarity, we list the adopted notations throughout this paper in  Table \ref{table1}.

\section{System Model}\label{sec:electromagnetic MIMO  model description}
\begin{figure}[htbp]
	\centering 
	\vspace{-0.35cm}
	\setlength{\abovecaptionskip}{0.5cm}
	\subfloat[Multi-user uplink transmission with HPAs equipped by both the receiver and the transmitters.]{
		\includegraphics[width=1\linewidth]{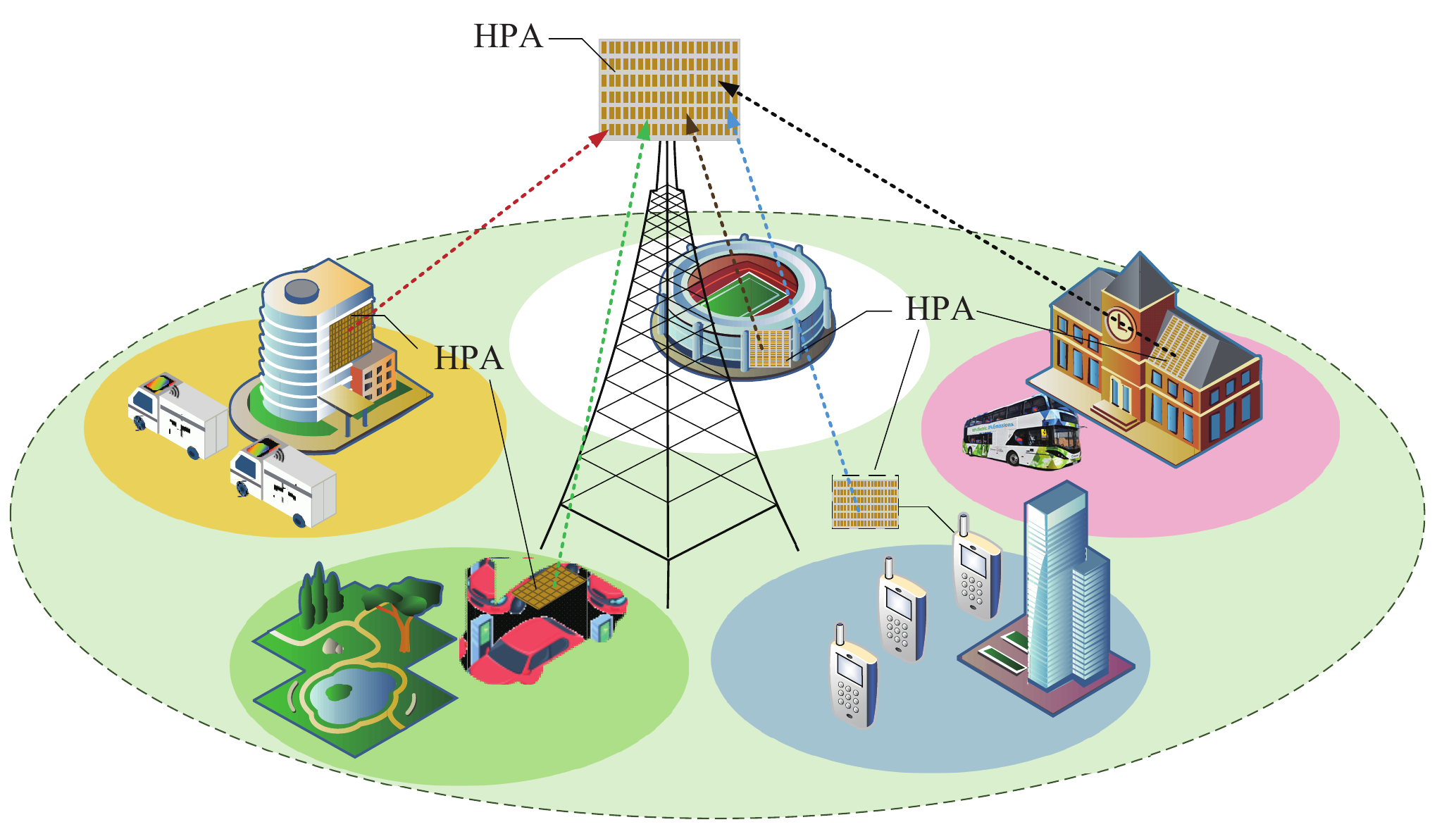}
		\label{Multiuser HMIMO}
	}\quad
	\vspace{-0.35cm}
	\subfloat[Communication  between the base station and one of the users.]{ 
		\includegraphics[width=1\linewidth]{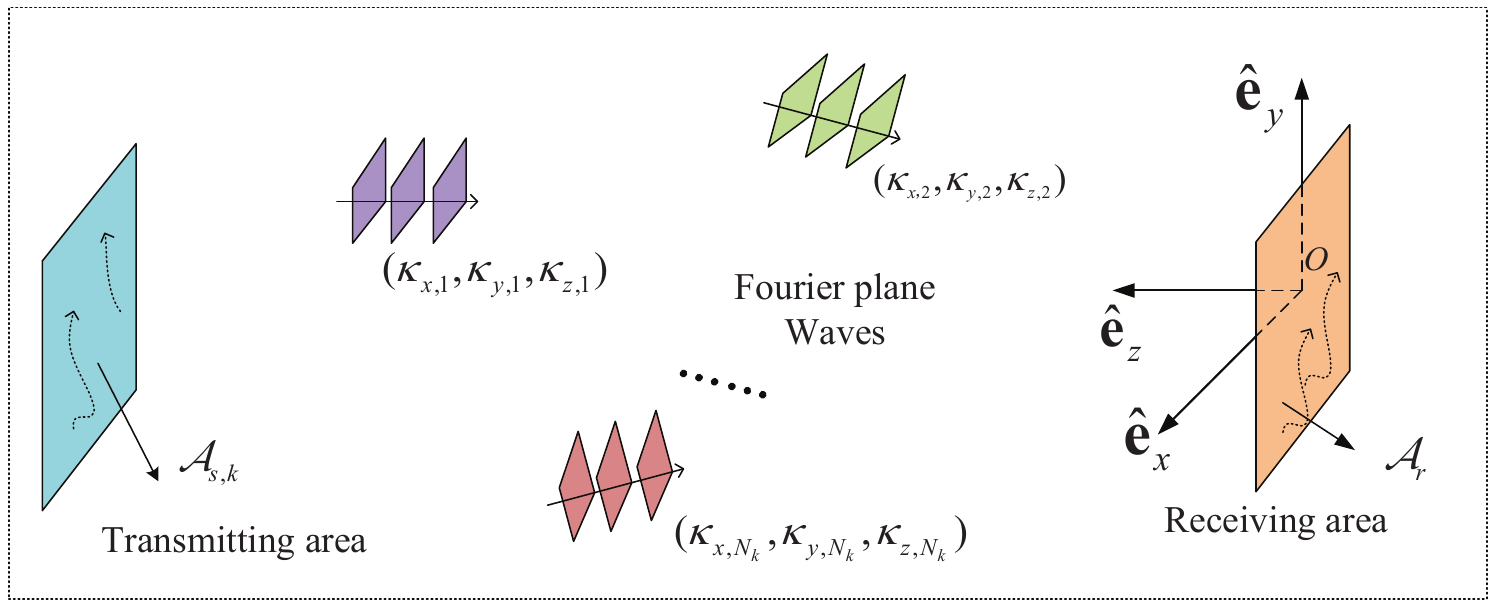}
		\label{Communication scenario}
	}
	\vspace{0.15cm}
	\caption{The HPAs-assisted multi-user uplink transmission.}
	\label{system setting}
\end{figure}

We consider the HPA-assisted uplink transmission between multiple users and the base station, which is shown in Fig. \subref*{Multiuser HMIMO}.
Taking into account the independence among channels from individual users to the base station in a multi-user uplink transmission system, our attention will be directed towards the channel between one of the users and the base station. We will illustrate the channel model from the EM perspective in Subsection \ref{subsec:EM channel}, followed by modeling the noise field in Subsection \ref{subsec: Noise field modelling}.
  
As illustrated in Fig. \subref*{Communication scenario}, we consider the free space transmission and  assume that the user and the base station are both equipped with rectangular HPAs. While an individual antenna has a fixed radiation pattern, antenna arrays are capable of changing their radiation patterns over time and frequency, for both transmission and reception.  With the progress of metamaterial technology, it is now possible to deploy more antennas, or even achieve approximate continuous antenna arrays, in limited space. Therefore, we consider the scenarios where HPAs are also deployed at the user end.
The receiving HPA, denoted as $\mathcal{A}_r$, is located on the $x$-$y$ coordinate plane, with its center at the origin $(0,0,0)$.
The coordinate ranges of point $\br$ on $\cA_{r}$ are  defined as follows,
\begin{align}
	\cA_{r}: \{(r_x,r_y,r_z):|r_x|\leq \frac{R_x}{2}, |r_y|\leq \frac{R_y}{2}, r_z =0 \}.
\end{align}

Because we consider multiple users with arbitrary spatial distributions, it implies that their HPAs cannot always remain parallel to the receiving  HPA. Therefore, to obtain the projection lengths of each user's HPA along the $x$ and $y$ axes in the coordinate system where the receiving HPA is located, we first perform a unitary transformation $\cU$ on each user's HPA.  Assuming the spatial position of a user's HPA is denoted by $\cA_s$, we use $\bs\triangleq(s_x,s_y,s_z)$ in the Cartesian coordinate system to denote the position of point  $\bs \in \cA_s$. $\cU \cA_{s} \triangleq \{\bU \bs : \bs \in \cA_{s}\}$ represents the plane parallel to the $x$-$y$ coordinate plane, which is  obtained by applying an orthogonal transformation $\cU$ with $\bU$ being the $3 \times 3$ real unitary matrix. Let $\bs' \triangleq (s_x',s_y',s_z') = \bU\bs$ and let $S_x$ and $S_y$ be the sizes of the projections of $\cU\cA_s$ along $x$ and $y$ axes, respectively, which are both non-zeros. 
Then, the ranges of $s_x'$, $s_y'$ and $s_z'$ can be expressed as follows
\begin{subequations}\label{UAs}
	\begin{align}
		&s_x': s_{x,0} \leq s_x' \leq s_{x,0} + S_x,\\
		&s_y': s_{y,0} \leq s_y' \leq s_{y,0} + S_y,\\
		&s_z': s_z' = s_{z,0}, \label{ranges of s_z}
	\end{align}
\end{subequations}
where $s_{x,0}$ and $s_{y,0}$ are the smallest coordinates of $s_x'$ and $s_y'$, respectively.  

For notation convenience, we use notions of $\bs$ and $\bs'$ in a 2D plane parallel to the $x$-$y$ plane and a 3D plane abusively.
It is necessary to point out that due to the different positions of users in the space, $\bU$ is user-dependent.




\subsection{Modeling of EM Channel}\label{subsec:EM channel}
We first illustrate the channel model with the help of the EM field equations.  Consider the case where an arbitrary current density $\bj(\mathbf{s},t) = \real{\bj(\bs)e^{-\jmath \omega t}}$ can be generated at any position $\bs \in \cA_{s}$  and time $t$,  with $\omega$ being the angular frequency of the current \cite{sanguinettiWavenumberDivisionMultiplexingLineofSight2023}. Simplifying the communication system to operate in narrowband, aligning with the commonly employed time-harmonic assumption in EM analysis, we can disregard the time-related component $e^{-\jmath \omega t}$ \cite{d2022cramer}. We focus on the time-independent current density, denoted as $\bj(\bs)$ \cite{10005192}, which can be expressed as follows
\begin{align}\label{current density function}
	\bj(\mathbf{s})=j_x(\mathbf{s}) \hat{\be}_x +j_y(\mathbf{s})\hat{\be}_y+ j_z(\mathbf{s})\hat{\be}_z, 
\end{align}
where  $\hat{\be}_x$, $\hat{\be}_y$ and $\hat{\be}_z$ are unit vectors in the three orthogonal directions $x$, $y$, and $z$ in the Cartesian coordinate system, respectively.
$j_x(\mathbf{s}), j_y(\mathbf{s})$ and $j_z(\mathbf{s})$ represent the scalar current density functions in the $x$, $y$, and $z$ directions, respectively. 

According to \cite[Eq. (1.3.50)]{chew1999waves}, the relationship between the current density $\bj(\bs)$ in the source region $\cA_s$ and the excited electric field $\be(\br)$ in the receiving region $\cA_r$ is 
\begin{align}\label{solution to Helmholtz equation}
	\be(\mathbf{r})=\int_{\mathcal{A}_s}  \mathbf{G}(\mathbf{r},\mathbf{s})\bj(\mathbf{s}) \mathrm{d} \mathbf{s},
\end{align}
where $\bs$ and $\br$ represent the points locating in $\cA_s$ and $\cA_r$, respectively.
$\mathbf{G}(\mathbf{r},\mathbf{s})$ is the dyadic Green's function in free space \cite{WaveTheoryofInformation}.  Since we only consider the radiation field (including Fresnel region and Fraunhofer region), the higher order terms of the Green's function with respect to the distance can be disregarded, keeping only the lower order terms \cite{bjornson2020power}, i.e.,
\begin{align}\label{Green}
	\bG(\mathbf{r},\mathbf{s}) = &-\frac{\jmath \eta \exp \left(-\jmath k_0 \|\mathbf{p}\|\right)}{2 \lambda \|\mathbf{p}\|}\left(\mathbf{I}-\hat{\bp} \cdot \hat{\bp}^H\right),
\end{align}
where $\bp = \br - \bs$ and $\hat{\mathbf{p}}=\mathbf{p}/\|\mathbf{p}\|$. $\kappa_0 = \omega /c $ is the free space wavenumber, where $c = 3\times 10^8 \: \mathrm{m/s}$  is  the speed of light in free space.  $\eta = \mu_0 c = 120\pi \: \Omega$  is the free space wave impedance with $\mu_0 = 4\pi \times 10^{-7} \: \mathrm{H/m}$. This approximation in \eqref{Green} is tight when the transmitter is beyond the reactive near-field of the receiving antenna \cite{bjornson2020power}.

From a signal processing perspective, it is worth noting that $\mathbf{G}(\br,\bs)$ can be interpreted as the 3D impulse response of the propagation medium, connecting the surface currents at point $\mathbf{s}$ with the induced fields at point $\mathbf{r}$ \cite{mikkiAntennaCurrentGreen2013}.
For homogeneous media, the dyadic Green's function $\bG(\br,\bs)$ depends  on the distance between the source point and the receiving point as well as on their individual positions.
The representation will be subsequently streamlined through the utilization of Fourier space series.

\subsection{Modeling of Noise Field}\label{subsec: Noise field modelling}
Regarding the modeling of the noise field, we have to consider not only the hardware noise at the receiver, but also the electric field excited by the current outside the source domain.
We denote $\bz(\br) = \bz^{\remi}(\br) + \bz^{\rhw}(\br)$ as the noise field function with $\br \in \cA_r$. The first part represented as $\bz^{\remi}(\mathbf{r})$ stems from the electric field excited by currents beyond the source region $\cA_s$, which we term EM interference. The second component, denoted as $\bz^{\rhw}(\mathbf{r})$, comprises the hardware noise originating from the receiving device and is independent of the first component $\bz^{\remi}(\br)$. 

Concerning the EM interference $\bz^{\remi}(\mathbf{r})$, it is assumed to follow an isotropic distribution \cite{sanguinettiWavenumberDivisionMultiplexingLineofSight2023}. This implies that waves may impinge from any elevation angle $\theta_r$ and any azimuth angle $\phi_r$. 
The corresponding function of $\bz^{\remi}(\br)$ is  \cite{sanguinettiWavenumberDivisionMultiplexingLineofSight2023},
\begin{align}\label{noise field}
	\bz^{\remi}(\mathbf{r}) = \int_{-\pi}^{\pi}\int_{0}^{\pi} \mathbf{a} (\theta_r,\phi_r) e^{\jmath \boldsymbol{\kappa}^{T}(\theta_r,\phi_r) \mathbf{r}}  \mathrm{d} \theta_r \mathrm{d} \phi_r , \quad \br \in \cA_r,
\end{align}
where $\bkappa(\theta_r,\phi_r)$ is the wave vector with $\theta_r \in [0,\pi]$ and $\phi_r \in [-\pi,\pi]$ :
\begin{align}\label{kappa_r}
	\boldsymbol{\kappa} (\theta_r, \phi_r) = \frac{2\pi}{\lambda} [\sin \theta_r \cos \phi_r, \sin \theta_r \sin \phi_r , \cos \theta_r]^{T},
\end{align}
and $\mathbf{a} (\theta_r,\phi_r)$ is a  random process that obeys a zero-mean complex-valued Gaussian distribution. In fact, we can regard $\mathbf{a} (\theta_r,\phi_r)$ as a plane wave function that can be incident from any angle $(\theta_r,\phi_r)$ in the space \cite{jensenCapacityContinuousSpaceElectromagnetic2008}.

 Due to the transverse wave nature of EM waves \cite{wanMutualInformationElectromagnetic2023}, the projection of $\mathbf{a} (\theta_r,\phi_r)$ in the direction of wave propagation should be $0$, i.e., $\mathbf{a}^{H} (\theta_r,\phi_r) \hat{\boldsymbol{\kappa}}(\theta_r, \phi_r) = 0$, where $\hat{\boldsymbol{\kappa}}(\theta_r, \phi_r) = \boldsymbol{\kappa}(\theta_r, \phi_r)/\|\boldsymbol{\kappa}(\theta_r, \phi_r)\|$ is the normalized wave vector. The autocorrelation function  of $\mathbf{a} (\theta_r,\phi_r) $ is given in \eqref{autocorrelation of a} at the bottom  of the next page, where $\sigma_{\remi}^2$ is the power angular density. 
\begin{figure*}[hb]
		\hrulefill
	\begin{subequations}\label{autocorrelation of a}
		\begin{align}
			\mathbb{E}\left[ \mathbf{a}(\theta_r,\phi_r)\mathbf{a}^{H}(\theta_{r^{\prime}},\phi_{r^{\prime}}) \right] = &  \quad \!\!\!\!
			\mathbb{E}\left[\mathbf{a}(\theta_r,\phi_r)\mathbf{a}^{H}(\theta_r,\phi_r)- 
			(\mathbf{a}^{H} (\theta_r,\phi_r) \hat{\boldsymbol{\kappa}} \cdot \mathbf{I})(\mathbf{a}^{H} (\theta_r,\phi_r) \hat{\boldsymbol{\kappa}} \cdot \mathbf{I} )^H   \right]  \\
			= & \quad  \!\!\!\!\sigma_{\remi}^2 f(\theta_r, \phi_r) (\mathbf{I}-\hat{\boldsymbol{\kappa}}\hat{\boldsymbol{\kappa}}^{T})  \delta(\theta_r-\theta_{r^{\prime}}) \delta(\phi_r-\phi_{r^{\prime}}).
		\end{align}
	\end{subequations}
\end{figure*}
Under the condition that the medium is isotropic, $f(\theta_r, \phi_r) = \sin \theta_r/4\pi$ \cite{sanguinettiWavenumberDivisionMultiplexingLineofSight2023}. As a result, the autocorrelation $\boldsymbol{\rho}(\br)$ of noise field $\bz^{\remi}(\mathbf{r}) $ is given in \eqref{autocorrelation of noise} at the bottom of next page, where $\mathbf{I}-\hat{\boldsymbol{\kappa}}(\theta_r, \phi_r) \hat{\boldsymbol{\kappa}}^{T}(\theta_r, \phi_r) $ is a $3 \times 3$ rank-2 matrix and constrains the oscillation direction of the radiated field to be perpendicular to its propagation direction. From \eqref{autocorrelation of noise}, it can be seen that the covariance matrix of the EM interference is not diagonal and its elements are correlated, i.e.,  the EM interference is spatially colored.
\begin{figure*}[hb]
	\hrulefill
	\begin{subequations}\label{autocorrelation of noise}
		\begin{align}
			\boldsymbol{\rho}(\br) = \mathbb{E}\left\{\bz^{\remi}\left(\mathbf{r}+\mathbf{r}^{\prime}\right) [\bz^{\remi}\left(\mathbf{r}^{\prime}\right)]^{H}\right\}  
			= & \int_{-\pi}^\pi \int_0^\pi \int_{-\pi}^\pi \int_0^\pi \mathbb{E}\left[\mathbf{a}(\theta, \phi) \mathbf{a}^{H}\left(\theta^{\prime}, \phi^{\prime}\right)\right] e^{\jmath \boldsymbol{\kappa}^{T} \cdot\left(\mathbf{r}+\mathbf{r}^{\prime}\right)} \cdot 
			e^{-\jmath \boldsymbol{\kappa}^{T} \cdot \mathbf{r}^{\prime}} 
			\mathrm{d} \theta  \mathrm{d} \phi  \mathrm{d} \theta^{\prime} \mathrm{d} \phi^{\prime}   \\ 
			= & \int_{-\pi}^\pi \int_0^\pi \sigma_{\remi}^2 f(\theta, \phi)\left(\mathbf{I}-\hat{\boldsymbol{\kappa}} \hat{\boldsymbol{\kappa}}^{T}\right) e^{\jmath \boldsymbol{\kappa}^{T} \cdot \mathbf{r}}  \mathrm{d} \theta  \mathrm{d} \phi .
		\end{align}
\end{subequations}
\end{figure*}

In comparison to EM interference $\bz^{\remi}(\mathbf{r})$, the modeling of  hardware noise, represented as $\bz^{\rhw}(\mathbf{r})$, is relatively straightforward. It can be modeled as  a spatially uncorrelated zero-mean
complex Gaussian process,  with correlation function  given as follows,
\begin{align}
	\mathbb{E}[ \bz^{\rhw}(\mathbf{r}) \left(\bz^{\rhw}(\mathbf{r}^{\prime}) \right)^H] = \frac{N_0}{2} \mathbf{I}_{3} \delta(\mathbf{r}-\mathbf{r}^{\prime}),
\end{align}
where $N_0$ denotes the power angular density of $\bz^{\rhw}(\mathbf{r})$ \cite{sanguinettiWavenumberDivisionMultiplexingLineofSight2023}. $\delta(\br-\br^\prime)$ is Dirac delta function, which is 1 when $\br =\br^\prime$ and 0 otherwise.
Note that  $\bz(\br)$ becomes a spatially colored noise model because EM interference is taken into account.

\section{Fourier Space Series Expansion of Current Density Function and Field Functions}\label{sec:Fourier-PWF}
Given that the  current density function $\bj(\bs)$, the excited electric field $\be(\br)$ and the noise field function $\bz(\br)$ are all continuous physical quantities, they have the potential to incur a prohibitively high computational burden when we analyze or optimize the performance of the system since continuous physical quantities mean that infinite dimensional parameters are involved.
To render them more tractable, a natural approach is to project them into a space spanned by a complete orthogonal function set and derive approximations within a finite-dimensional space through reasonable truncation. 
We will clarify the expansions of the source current density function, the excited electric field function and the noise field function  in Subsections \ref{subsec:current density function}, \ref{subsec: Excited Electric Field Function} and \ref{sec:noise field expansion}, respectively.

The authors  in \cite{millerCommunicatingWavesVolumes2000} 
introduced the concept of ``\textit{communication modes}'', which are essentially pairs of functions. Each pair comprises one function for the transmitting volume and another for the receiving volume. Importantly, there exists a direct correspondence between each pair of functions across the two volumes, facilitated by a multiplicative coupling coefficient.
In the sense of maximizing the sum of the strengths of the coupling coefficients, the optimal basis functions can be selected as the eigenfunctions of the channel operator, which is obtained by performing Hilbert–Schmidt decomposition \cite{millerCommunicatingWavesVolumes2000}. 

However, it is important to note that solving the eigenfunction problem, which defines the optimal sets of basis functions and coupling coefficients at both the transmitter and the receiver, has high computational complexity.
Therefore,  basis functions such as Fourier space series, with suboptimal performance but much lower complexity,  are more desirable  for practical applications. In fact, the outward fields can be efficiently computed via  discrete	Fourier transform (DFT). Additionally, plane waves are origin-independent, implying that we solely need to specify their respective strength.
Based on this consideration, the Fourier space series will be chosen as the basis functions of the input and output signal subspaces in the following study\cite{bjornson6GMIMOMassive2024}.

\subsection{Current Density Function}\label{subsec:current density function}
To facilitate the representation of the integration interval,  we can change the integration from $\int_{\cA_{s}} \ds$ to $\int_{\cU\cA_{s}} \det(\bJ_\cU)\ds'$. $\bJ_\cU$ is the Jacobian matrix and since $\cU$ is the orthogonal transformation,  $\det(\bJ_\cU) = \pm 1$ \cite{ahmed2012orthogonal}.
In this subsection, we first expand the continuous current density function $\bj(\bs)$ in terms of two-dimensional Fourier space series, i.e., 
\begin{align}
	\bj(\bs)= & \bj (\bU^T \bs') \notag \\
	= &\sum_{\bn} \bxi_{\bn} \phi_{\bn} (\bs') , \quad  \bs' \in \cU\cA_s,\notag \\
	=&\sum_{\bn} \bxi_{\bn} \phi_{\bn} (\bU \bs) , \quad  \bs \in \cA_s,
\end{align}
where the coefficient $\bxi_{\bn}$ is
\begin{subequations}
	\begin{align}\label{xi_nm}
		\bxi_{\bn} = & \frac{1}{S_x S_y}\int_{\cU\mathcal{A}_s} \bj(\bU^T \bs') \phi_{\bn}^{*}(\bs') \ds' \\
		= & \frac{1}{S_x S_y}\int_{\mathcal{A}_s} \bj(\bs) \phi_{\bn}^{*}(\bU\bs) \det(\bJ_{\cU})\ds.
	\end{align}
\end{subequations}

The basis function $\phi_{\bn}(\bs'),\bs' \in \cU\cA_s$, is \cite{sanguinettiWavenumberDivisionMultiplexingLineofSight2023},
\begin{align}\label{phi_n}
	\phi_{\bn} (\bs') &= \frac{1}{\sqrt{S_x S_y}}e^{\jmath \bkappa_\bn \bs'} \notag \\
	&=\frac{1}{\sqrt{S_x S_y}}e^{\jmath \left(\kappa_{x,n_x}s_x' +\kappa_{y,n_y}s_y' \right)}.
\end{align}
In \eqref{phi_n}, $\bn \triangleq (n_x,n_y)$ is the two dimensional index for the basis function $\phi_{\bn} (\bs')$ in the input space while $n_x$ and $n_y$ are integers.
$\bkappa_\bn  \triangleq \left(\kappa_{x,n_x},\kappa_{x,n_y}\right)$, which is labeled by the subscript $\bn$, represents the sampling wavenumber,  with $\kappa_{x,n_x}$ and $\kappa_{y,n_y}$ given as follows
\begin{subequations}\label{kappa_n}
		\begin{align}
			\kappa_{x,n_x} = &\frac{2\pi}{S_x}n_x,  \\
			\kappa_{y,n_y} = &\frac{2\pi}{S_y}n_y.
		\end{align}
\end{subequations}


Note that $\bkappa_\bn$ varies with $S_x$, $S_y$ and $\bn$. Since we are considering propagating waves rather than evanescent waves, the sampling wavenumber $\bkappa_\bn$ needs to satisfy the following condition \cite{Li:09,zhangPatternDivisionMultiplexingMultiUser2023,10005192},
	\begin{align}\label{ranges of kappa}
		\kappa_{x,n_x}^2 +  \kappa_{x,n_y}^2  \leq \kappa_0^2.
	\end{align}
Substituting \eqref{kappa_n} into \eqref{ranges of kappa}, the ranges of $n_x$ and $n_y$ are 
\begin{subequations}\label{range_n}
		\begin{align}
			|n_x| \leq \lceil  \frac{S_x}{\lambda}\rceil, \\
			|n_y| \leq \lceil  \frac{S_y}{\lambda}\rceil.
		\end{align}
\end{subequations}

It can be seen that due to the constraints of \eqref{ranges of kappa}, we can take advantage of the finite bandwidth property of the wavenumber domain to truncate the finite items in the expansion of the current density function $\bj(\bs)$, which suggests that processing signals in the wavenumber domain will avoid the high computational complexity associated with the infinite dimensional parameters.

\subsection{Excited Electric Field Function}\label{subsec: Excited Electric Field Function}
Since $r_z = 0$, we will omit it in the following and let $\br\triangleq(r_x,r_y)$  as mentioned in the paragraph below \eqref{UAs}. 
Similar to the approximation procedure for the current density function $\bj(\bs)$, we project the excited electric field $\be(\br)$ into finite-dimensional space spanned by the following basis functions $\psi_{\bbm}(\br),\br \in \cA_r$,
\begin{align}\label{psi_m}
	\psi_{\bbm}(\br) &= \frac{1}{\sqrt{R_x R_y}}e^{\jmath \bkappa_\bbm \br} \notag \\
	&= \frac{1}{\sqrt{R_x R_y}}e^{\jmath\left(\kappa_{x,m_x}r_x  + \kappa_{y,m_y}r_y \right)}, 
\end{align}
where $\bbm \triangleq (m_x,m_y)$ and the elements of $\bkappa_\bbm$ are similarly defined in \eqref{kappa_n}, i.e., $\bkappa_\bbm \triangleq (\kappa_{x,m_x},\kappa_{y,m_y})\triangleq (\frac{2\pi}{R_x}m_x,\frac{2\pi}{R_y}m_y)$.  Since $\kappa_{x,m_x}$ and $\kappa_{y,m_y}$ also need to satisfy $\kappa_{x,m_x}^2+\kappa_{y,m_y}^2 \leq \kappa_0^2$, the ranges of $m_x$ and $m_y$ are 
\begin{subequations}
	\begin{align}
		|m_x|\leq \lceil \frac{R_x}{\lambda}\rceil,\\
		|m_y|\leq \lceil\frac{R_y}{\lambda}\rceil.
	\end{align}
\end{subequations}

Under the basis function $\psi_\bbm(\br)$, the projection of $\be(\br)$ is given by
\begin{subequations}\label{discrete E}
	\begin{align}
		\be_{\bbm} = &\int_{\mathcal{A}_r} \psi_{\bbm}^{*}(\mathbf{r}) \be(\mathbf{r}) \rd \mathbf{r} \\
		=& \int_{\mathcal{A}_r}  \psi_{\bbm}^{*}(\mathbf{r}) \int_{\cA_s} \bG(\br,\bs)\bj(\bs) \ds  \dr \\
		=&\int_{\mathcal{A}_r}  \psi_{\bbm}^{*}(\mathbf{r}) \int_{\cA_s} \bG(\br,\bs)\sum_{\bn} \bxi_{\bn} \phi_{\bn} (\bU \bs)\ds  \dr \\
		=&\sum_{\bn}  \left(\int_{\mathcal{A}_r} \int_{\cA_s} \psi^{*}_{\bbm}(\br) \bG(\br,\bs) \phi_{\bn}(\bU\bs)  \ds \dr \right) \bxi_{\bn}\label{H_definition}\\
		\triangleq & \sum_{\bn} \bH_{\bbm,\bn} \bxi_{\bn}. \label{Hmn}
	\end{align}
\end{subequations}
It is important to recognize that despite the resemblance of the form of $\bH_{\bbm,\bn}$ to a channel matrix, it actually conveys the level of coupling between the transmitting mode $\bkappa_\bn$ and the receiving mode $\bkappa_\bbm$. Consequently, $\bH_{\bbm,\bn}$ is defined as the coupling coefficient \cite{sanguinettiWavenumberDivisionMultiplexingLineofSight2023}. 
A high level of coupling implies that the EM waves predominantly excited by the current density function are confined to the region where the receiving plane is located, while a low level of coupling implies that a significant portion of the wave dissipates in other spatial locations and cannot be captured by the receiver\cite{dardariHolographicCommunicationUsing2021}. 
The advantage of EM modeling is that it provides a description of the channel when the array tends to be continuous since traditional channel models are not applicable when antenna element spacing is tiny.

To clarify the connection between $\bH_{\bbm,\bn}$ and $\bG(\mathbf{r},\bs)$, we introduce the notation $\overline{\bG}(\bkappa)$, which represents the outcome of the spatial Fourier transformation applied to $\bG(\br,\bs)$. Then, $\bG(\br,\bs)$ can be re-expressed as follows,
\begin{align}\label{re_G}
	\bG(\br,\bs) = \bG(\bp) = \frac{1}{(2\pi)^2} \int_{\cR^2} \overline{\bG}(\bkappa) e^{\jmath \bkappa \bp} \rd \bkappa,
\end{align}
where $\bkappa \triangleq (\kappa_x,\kappa_y)$ is the wavenumber and $\bp = \br - \bs$.
\begin{figure}[htbp]
	\centering 
	\vspace{-0.35cm}
	\setlength{\abovecaptionskip}{0.5cm}
	\subfloat[Two-dimensional schematic of the  wavenumber-domain Green's function $\overline{\bG}(\kappa_x,\kappa_y)$.]{
		\includegraphics[width=1\linewidth]{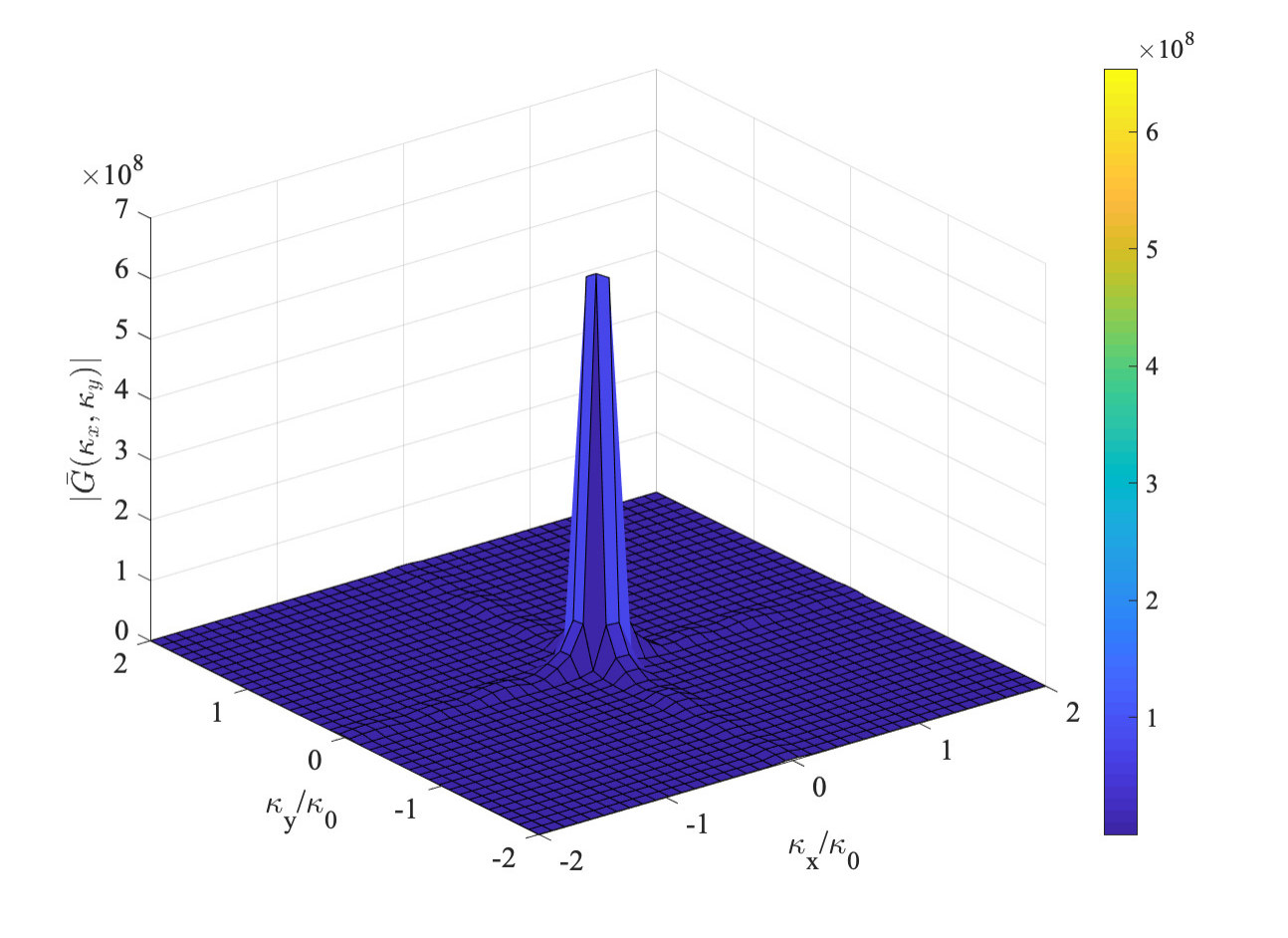}
		\label{wavenumber G}
	}\quad
	\vspace{2mm}
	\subfloat[Cross section of $\overline{\bG}(\kappa_x,\kappa_y)$ along $\kappa_y$ direction.]{ 
		\includegraphics[width=0.9\linewidth]{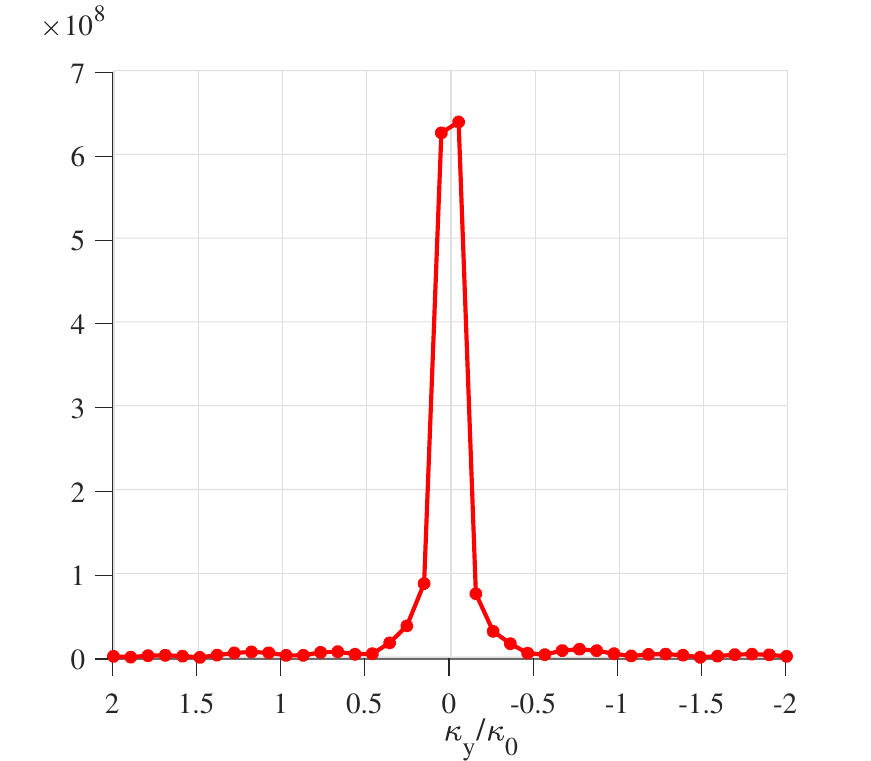}
		\label{cross section of G}
	}
	\caption{Overview of Green's function in the wavenumber domain.}
	\label{Wavenumber Green's function}
\end{figure}
The depiction of $\overline{\bG}(\bkappa)$ is presented in \figref{Wavenumber Green's function}. It is evident that the spatial bandwidth of $\overline{\bG}(\bkappa)$ falls within the range of $[-\kappa_0, \kappa_0]$. This observation indicates that a finite number of terms can be effectively truncated when expanding the source current density function and the field functions using the Fourier space series.

Substituting \eqref{re_G} into \eqref{H_definition}, $\bH_{\bbm,\bn}$ can be re-expressed as \eqref{H_mn} shows, which is on the top of next page.
\begin{figure*}[ht]
\begin{subequations}\label{H_mn}
\begin{align}
		\bH_{\bbm,\bn} = & \int_{\mathcal{A}_r} \int_{\cA_s} \psi^{*}_{\bbm}(\br) \bG(\br,\bs) \phi_{\bn}(\bU\bs)  \ds \dr \\
			=& \int_{\mathcal{A}_r} \int_{\cA_s} \psi^{*}_{\bbm}(\br) [\frac{1}{(2\pi)^2} \int_{\cR^2} \overline{\bG}(\bkappa) e^{\jmath \bkappa \bp} \rd \bkappa] \phi_{\bn}(\bU\bs)  \ds \dr \\
			=&\frac{1}{(2\pi)^2} \int_{\cR^2}  \int_{\mathcal{A}_r} \int_{\cA_s} \psi^{*}_{\bbm}(\br) \overline{\bG}(\bkappa) e^{\jmath \left(\kappa_x(r_x-s_x)+\kappa_y(r_y-s_y)\right)}  \phi_{\bn}(\bU\bs)  \ds \dr    \rd \bkappa  \\
			=&\frac{1}{(2\pi)^2} \int_{\cR^2}   \int_{\mathcal{A}_r} \int_{\cA_s} \psi^{*}_{\bbm}(\br)  e^{\jmath \left(\kappa_x r_x+\kappa_y r_y\right)} \overline{\bG}(\bkappa) e^{-\jmath \left(\kappa_x s_x+\kappa_y s_y\right)}\phi_{\bn}(\bU\bs)  \ds \dr    \rd \bkappa \\
			=& \frac{1}{(2\pi)^2} \int_{\cR^2}   \left[ \int_{\mathcal{A}_r} \psi^{*}_{\bbm}(\br)  e^{\jmath \left(\kappa_x r_x+\kappa_y r_y\right)}\dr\right] \overline{\bG}(\bkappa) \left[\int_{\cA_s} e^{-\jmath \left(\kappa_x s_x+\kappa_y s_y\right)}\phi_{\bn}(\bU\bs)  \ds \right]     \rd \bkappa \\
		= &\frac{1}{(2\pi)^2}\int_{\cR^2} \Psi_{\bbm}^{*}(\bkappa)  \overline{\bG}(\bkappa)  \Phi_{\bn}(\bkappa)  \rd \bkappa \label{H_m,n}.
	\end{align}
\end{subequations}
\hrulefill
\end{figure*}
In \eqref{H_m,n}, $\Psi_{\bbm}(\bkappa)$ and $\Phi_{\bn}(\bkappa)$ are the analog spatial Fourier transformations of $\psi_{\bbm}(\br)$ and $\phi_{\bn}(\bU\bs)$ that are 0 outside $\cA_s$ and $\cA_r$, respectively, given as follows,
\begin{subequations}\label{sinc}
\begin{align}
   	\Phi_{\bn}(\bkappa) = & \int_{\cA_{s}} \phi_{\bn}(\bU\bs)e^{-\jmath\bkappa \bs}\ds \\
   	= &\int_{\cU\cA_{s}} \phi_{\bn}(\bs')e^{-\jmath \bkappa \bU^T \bs'} \det(\bJ_{\cU})^{-1}\ds', \label{Phi_kappa}\\
	\Psi_{\bbm}(\bkappa) = & \int_{\cA_{r}} \psi_{\bbm}(\br)e^{-\jmath\bkappa \br}\dr \label{Psi_kappa} \\
	= & \delta(\bkappa - \bkappa_\bbm), \quad R_x, R_y \to \infty \label{Psi_infinity}.
\end{align}
\end{subequations}

After disregarding irrelevant constant terms, \eqref{Phi_kappa} corresponds to a sinc function,  and it attains its maximum value when $\bkappa  = \bkappa_\bn \bU $, while \eqref{Psi_kappa} also corresponds to a sinc function and it attains its maximum value when $\bkappa  =  \bkappa_\bbm$. Because $\overline{\bG}(\bkappa)$ denotes the wavenumber domain representation of the spatial channel response $\bG(\br,\bs)$, $\bH_{\bbm,\bn}$ can be viewed as the sampling of $\overline{\bG}(\bkappa)$, which indicates the degree of coupling between transmitting mode $\bkappa_\bn$  and the receiving mode $\bkappa_\bbm$. 
In particular, $\Psi_{\bbm}(\bkappa)$ can be approximated as a Dirac delta function when $R_x$ and $R_y$ tend to be infinity as \eqref{Psi_infinity} shows. In this case, the coupling matrix $\bH_{\bbm,\bn} = \overline{\bG}(\bkappa_\bbm)\Phi_{\bn}(\bkappa_\bbm)$  indicates that when a wave is transmitted for mode $\bkappa_\bn$, its coupling to mode $\bkappa_\bbm \bU$ is at its maximum level. 


\subsection{Noise Field Function}\label{sec:noise field expansion}
After considering the projection of the excited electric field $\be(\mathbf{r})$ into the subspace spanned by the given basis functions $\psi_\bbm(\br)$, the projection of the noise field $\bz(\br)$ also needs to be considered.
The projection of noise under basis function $\psi_{\bbm}(\br)$ is $\bz_{\bbm} = \bz_{\bbm}^{\remi} + \bz_{\bbm}^{\rhw} $,  where
\begin{subequations}\label{z}
	\begin{align}
		\bz_{\bbm}^{\remi} = \int_{\mathcal{A}_r} \psi^{*}_{\bbm}(\mathbf{r}) \bz^{\remi}(\mathbf{r}) \rd \mathbf{r}, \\
		\bz_{\bbm}^{\rhw} = \int_{\mathcal{A}_r} \psi^{*}_{\bbm}(\mathbf{r}) \bz^{\rhw}(\mathbf{r}) \rd \mathbf{r},
	\end{align}
\end{subequations}
are mutually independent, and $\bbm \triangleq  (m_x,m_y)$. 

To make the representation clear, we denote the ranges of $m_x$ and $m_y$ as $[-M_x,M_x]$ and $[-M_y,M_y]$, respectively.  Thus, the range of $\bbm$  can be denoted as $\mathcal{M} \triangleq \{\bbm_j|j \in \{ 1,2,\dots,M \}\}$, where $M$ is
the dimension of the output space and can be written as $M = (2M_x+1)(2M_y+1)$. 
Without causing any confusion, specify that the range of $\bbm_i \in \cM$, is obtained by cyclically changing the values of $m_x$ and $m_y$. Let the first element in $\cM$ be $\bbm_1 = (-M_x,-M_y)$ and the last element in $\cM$ be $\bbm_M = (M_x,M_y)$.  It is then possible to arrange the noise projection values $\bz_\bbm$ under each receiving basis function $\psi_{\bbm}(\br)$ into the following super vector $\bz = [\bz_{\bbm_1},\bz_{\bbm_2},\dots,\bz_{\bbm_M}]^T$, where each element is a $3 \times 1$ dimensional vector. 

Going forward, we will denote the covariance matrix of $\bz$ as $\mathbf{R}_{\bz}$,  which is composed of two parts,
\begin{align}\label{covariance of colored noise}
	\mathbf{R}_{\mathbf{z}} = \sigma_{\mathrm{emi}}^{2} \mathbf{R}_{\mathbf{m}}^{\mathrm{emi}}+ \frac{N_0}{2}\mathbf{I},
\end{align}
where the $(j,i)$-th element of $\bR_{\bbm}^{\remi}$ is 
\begin{align}
	[\bR_{\bbm}^{\remi}]_{j,i} = \int_{\mathcal{A}_r} \int_{\mathcal{A}_r}  \psi_{\bbm_j}^{*}(\br)   \boldsymbol{\rho}(\br-\br')  \psi_{\bbm_i}(\br')\dr\dr',
\end{align}
and $\boldsymbol{\rho}(\br-\br')$ is given in \eqref{autocorrelation of noise}. 

At this point, we have illustrated the EM channel model and the colored noise model for the communication link between one of the users and the base station. By employing Fourier space basis functions, we have approximated the continuous models in a finite-dimensional space. 
Building upon the finite-dimensional models, the subsequent sections will delve into an analysis of how to maximize the system SE.

\section{Holographic Precoding for HPAs-Assisted Multi-user Uplink Transmission System }\label{sec: Current density function design}
In this section, we will investigate the HPAs-assisted multi-user uplink transmission. Specifically, in Subsection \ref{subsec: problem formulation}, we will formulate the problem of maximizing the system SE, and in Subsection \ref{subsec: Power allocation}, we will present an optimization algorithm for this problem and analyze the complexity as well as the convergence of the algorithm.

\subsection{Problem Formulation}\label{subsec: problem formulation}
We assume that there are $K$ users and let $\mathcal{K} \triangleq \{1,2,\dots,K\}$ represent the set of users. In Section \ref{sec:electromagnetic MIMO  model description} and Section \ref{sec:Fourier-PWF}, we have illustrated the channel model by taking the channel between one of the users and the base station as an example. 
Based on the model, we add subscript $k$ to the previous representations to distinguish the channels between different users and the base station below. 
The plane where the HPA of the $k$-th user is located is designated as $\cA_{s,k}$. 
Due to the possibility of HPAs of different users being located at arbitrary positions in the space, in order to represent the spatial positions of each HPA as well as their effective sizes on the $x$ and $y$ axes in the same coordinate system, we perform orthogonal transformations on the HPA of each user, enabling them to be described in a coordinate system parallel to the $x$-$y$ plane.
Let $\cU_k$ represent such an orthogonal transformation performed on $\cA_{s,k}$, i.e., $\bs' = \bU_k \bs, \forall \bs \in \cA_{s,k}$. $\bU_k$ is a $3\times 3$ dimensional real unitary matrix and is user-dependent.
The lengths of the projections of $\cU_k \cA_{s,k}$ along the $x$ and $y$ axes are denoted as $S_{k,x}$ and $S_{k,y}$, respectively. The positions and the sizes of the receiving HPA at the base station remain unchanged.

As the current density function $\bj(\bs)$ has components $j_x(\bs)$, $j_y(\bs)$ and $j_z(\bs)$ along three orthogonal directions $x$, $y$ and $z$, respectively, we assume that component of each direction can carry a symbol, and each user equipped with the HPA can transmit multiple data streams.
Then, the  synthesized current density function for the $k$-th user is given as follows,
\begin{align}
	\bx_k(\bs) = \sum_{\ell = 1}^{L_k} \bj_{k,\ell}(\bs) \odot \bs_{k,\ell},\quad k \in \cK,
\end{align}
where $\bs_{k,\ell} \in \bbC^{3\times 1}$ is the symbol transmitted by the $\ell$-th current density $\bj_{k,\ell}$ of the $k$-th user. Assume $\bs_{k,\ell} = [s_{k,\ell,x}, s_{k,\ell,y}, s_{k,\ell,z}]^T$ satisfies $\expect{\bs_{k,\ell} \bs_{k,\ell}^H} = \bI_3$ while $\expect{\bs_{k,\ell} \bs_{k,\ell'}^H} = \bzero, \forall \ell \neq \ell'$ and $\expect{\bs_{k,\ell} \bs_{k',\ell}^H} = \bzero, \forall k \neq k'$. $L_k$ is the number of the data streams transmitted by the $k$-th user.
$\bj_{k,\ell}(\bs)$ is assumed to be an arbitrarily current density function
(charge flux) of any external charges (not including any induced polarization currents), which is measured in units of $[\text{A}/\mathrm{m}^2]$.
Since each current density function $\bj_{k,\ell}(\bs)$ can be synthesized separately at the time of transmission, the  power constraint for the $k$-th user is
\begin{align}\label{power_constraint}
\int_{\cA_{s,k}} \expect{||\bx_k(\bs)||^2} \ds = \sum_{\ell = 1}^{L_k} \int_{\cA_{s,k}} ||\bj_{k,\ell}||^2  \ds \leq P_{k,\max}.
\end{align}

The received signal $\by(\br)$ is given as follows,
\begin{subequations}\label{y}
	\begin{align}
		\by(\br) &=  \sum_{k = 1}^{K}\be_k(\br) + \bz(\br)  , \\
		& = \sum_{k=1}^{K}   \int_{\cA_{s,k}} \bG(\br,\bs)  \bx_k(\bs)  \ds   + \bz(\br), \\
		& = \sum_{k=1}^{K} \sum_{\ell = 1}^{L_k}  \int_{\cA_{s,k}} \bG(\br,\bs)  \bj_{k,\ell}(\bs) \odot \bs_{k,\ell}  \ds   + \bz(\br), 
	\end{align}
\end{subequations}
where $\bz(\br)$ is the colored noise derived in Subsection \ref{subsec: Noise field modelling}, i.e., $\bz(\br) = \bz^{\remi}(\br) + \bz^{\rhw}(\br)$. 

In the given configuration, the expression for the system SE is 
\begin{align}
	R_{\rsum} = \log_2  \det\left( \bI_3 + \sum_{k = 1}^{K}\bp_k(\br) \bp_k^H(\br) \bR_{c,\bz}^{-1}(\br,\br') \right), 
\end{align}
where
\begin{align}
	\bp_k(\br) = \sum_{\ell = 1}^{L_k}  \int_{\cA_{s,k}} \bG(\br,\bs) \bj_{k,\ell}(\bs) \odot \bs_{k,\ell} \ds,
\end{align}
and
\begin{align}
		\bR_{c,\bz}(\br,\br') = \boldsymbol{\rho}(\br-\br') + \frac{N_0}{2}\bI_3\delta(\br-\br').
\end{align}

The sum-rate maximization problem can be formulated as follows,
\begin{subequations}\label{sum rate A}
	\begin{align}
		\cP_A: \quad \max_{\bj_{k,\ell}(\bs)}&  \quad R_\rsum \\
		\st&  \quad \sum_{\ell = 1}^{L_k}  \int_{\cA_{s,k}}||\bj_{k,\ell}(\bs)||^2 \ds \leq P_{k,\max},  \forall k \in \cK.
	\end{align}
\end{subequations}

The optimization variables $\{\bj_{k,\ell}(\bs)\}_{\forall k,\ell}$ in this problem are  continuous functions, making the optimization quite challenging. This is because it implies that computations may take place in an infinite-dimensional space, rendering conventional optimization methods no longer applicable. In the next subsection, we will transform this optimization problem into the optimization in a finite-dimensional space based on the expansions by Fourier space series.
\subsection{Optimization of Power Allocation Matrices}\label{subsec: Power allocation}
Since the impact of the transmitted current density function in the receiving region, for different symbols and users, can be obtained using the superposition theorem, we first analyze its discrete approximation in a finite-dimensional space for the $\ell$-th symbol-carrying function of the $k$-th user, denoted as $\mathbf{j}_{k,\ell}$.

Based on the analyses in Section \ref{sec:Fourier-PWF}, we project the current density functions $\bj_{k,\ell}(\bs) = \bj_{k,\ell}\left(\bU_k^T\bs'\right)$ into a subspace spanned by basis functions $\{\phi_{k,\bn}(\bs')\}_{\bn},\forall k$, i.e.,
\begin{align}\label{theta_k}
	\bj_{k,\ell}(\bs) = &\bj_{k,\ell}(\bU_k^T\bs') \notag \\
	=& \sum_{\bn\in \cN_k} \bxi_{k,\ell,\bn} \phi_{k,\bn}(\bs'), \quad \bs' \in \cU_k\cA_{s,k},\notag \\
	=&\sum_{\bn\in \cN_k} \bxi_{k,\ell,\bn} \phi_{k,\bn}(\bU_k \bs), \quad \bs \in \cA_{s,k},
\end{align}
where $\phi_{k,\bn}(\bs')$ is the basis function similarly defined in \eqref{phi_n}. 

The diverse positions of the HPA of each user result in different lengths of projections of $\cU_k\cA_{s,k}$ along the $x$ and $y$ axes, denoted as $S_{k,x}$ and $S_{k,y}$, respectively. Consequently, the values of $\kappa_{x,n_x}$ and $\kappa_{y,n_y}$ in \eqref{kappa_n}, defined by $S_{k,x}$ and $S_{k,y}$, respectively, will be associated with the parameter $k$.
$\bxi_{k,\ell,\bn}$ is the projection result of $\bj_{k,\ell}(\bU_k^T\bs')$ in the three orthogonal directions $x, y$ and $ z$. 
The range of $\bn$ for the $k$-th user is defined as $\mathcal{N}_k = [-N_{k,x},N_{k,x}]\times [-N_{k,y},N_{k,y}]$, 
where
\begin{subequations}
	\begin{align}
		&N_{k,x} = \lceil \frac{ S_{k,x}}{\lambda}\rceil,  \\
		&N_{k,y} = \lceil \frac{ S_{k,y}}{\lambda}\rceil.
	\end{align}
\end{subequations}
Thus, the input space dimension of the $k$-th user  can be denoted as $N_k = |\mathcal{N}_k| = (2N_{k,x}+1)(2N_{k,y}+1)$. 
For convenience of representation, we denote the first element of $\cN_k$ as $\bn_{1} = (-N_{k,x},-N_{k,y})$ and the last element as $\bn_{N_k} = (N_{k,x},N_{k,y})$. The other values $\bn_{i}$  in between are obtained by changing $n_x$ and $n_y$ sequentially.  In what follows, we will use $\bxi_{k,\ell,i}$ and $\phi_{k,\bn_{i}}$ instead of $\bxi_{k,\ell,\bn}$ and $\phi_{k,\bn}$ for a clearer description.

Since $\bx_{k,\ell}(\bs) \triangleq \bj_{k,\ell}(\bs) \odot \bs_{k,\ell} = \sum_{i = 1}^{N_k} \left(\bxi_{k,\ell,i}\odot \bs_{k,\ell}\right) \phi_{k,\bn_{i}}$, 
the power constraint in \eqref{power_constraint} can be reformulated into the following form according to Parseval's law,
\begin{align}\label{power constraint}
	\sum_{\ell = 1}^{L_k}\sum_{i = 1}^{N_k} \expect{||\bxi_{k,\ell,i} \odot \bs_{k,\ell}||^2 }
	=\sum_{\ell = 1}^{L_k} ||\bxi_{k,\ell}||^2 \leq P_{k,\max}, 
\end{align}
where $\bxi_{k,\ell}$ is the $3N_k \times 1$ super vector  with its $i$-th element being $\bxi_{k,\ell,i} \in \bbC^{3\times 1}$.
As a result of the basis function expansion, the continuous optimization variables $\{\bj_{k,\ell}(\bs)\}_{\forall k,\ell}$ become discrete expansion coefficients $\{\bxi_{k,\ell}\}_{\forall k,\ell}$.

Perform the Fourier space series expansion for the received signal by using the basis functions $\psi_{\bbm}(\br)$ in \eqref{psi_m}.
We still represent the range of $\bbm$ as $\cM$, 
which has been clarified in Subsection \ref{sec:noise field expansion}.
The projection of $\be_k(\br)$ along the $j$-th basis function $\psi_{\bbm_j}(\br)$ is
\begin{subequations}
    \begin{align}
    	&\be_{k,j} = \int_{\cA_r} \psi_{\bbm_j}^{*}(\br)\int_{\cA_{s,k}} \bG(\br,\bs)\bx_k(\bs)\ds \dr \\
    	&=\sum_{\ell = 1}^{L_k}\int_{\cA_r} \psi_{\bbm_j}^{*}(\br)\int_{\cA_{s,k}} \bG(\br,\bs)\bj_{k,\ell}\odot \bs_{k,\ell} \ds \dr \\
    	&= \sum_{\ell = 1}^{L_k} \sum_{i = 1}^{N_k} \bH_{k,j,i}  \bxi_{k,\ell,i} \odot \bs_{k,\ell} , \quad  j = 1,2,\dots,M,
    \end{align}
\end{subequations}
where $\bH_{k,j,i} $ is given by
\begin{align}\label{H_kji}
	&\bH_{k,j,i} = \int_{\mathcal{A}_r} \int_{\mathcal{A}_{s,k}} \psi^{*}_{\bbm_j}(\mathbf{r}) \bG(\br,\bs) \phi_{k,\bn_i}(\bU_k\bs) \ds \dr.
\end{align}


Let $\bH_{k} \in \bbC^{3M\times 3N_k}$ represent the coupling block matrix with its $(j,i)$-th block being $\bH_{k,j,i} \in \bbC^{3\times 3}$ and $\bXi_{k,\ell} = \mathbf{X}_{k,\ell}\bX_{k,\ell}^H$ with $\bX_{k,\ell}$ defined as follows
\begin{align}
\bX_{k,\ell} = \begin{bmatrix}
	\rdiag{\bxi_{k,\ell,1}} \\
	\rdiag{\bxi_{k,\ell,2}}\\
	\vdots\\
	\rdiag{\bxi_{k,\ell,N_k}}
\end{bmatrix} \in \bbC^{3N_k \times 3}.
\end{align}
The sum-rate maximization problem $\cP_{A}$ in \eqref{sum rate A} can be reformulated as 
\begin{subequations}\label{cP_tildeA}
	\begin{align}
		\mathcal{P}_{\tilde{A}}: \quad \max_{\boldsymbol{\Xi}_{k}}& \quad \tilde{R}_\mathrm{sum} \\
		\st& \quad  \sum_{\ell = 1}^{L_k} \trace{\boldsymbol{\Xi}_{k,\ell}}  \leq P_{k,\max},  \quad \forall k\in \cK,
	\end{align}
\end{subequations}
where $\sum_{\ell = 1}^{L_k} \trace{\bXi_{k,\ell}} = \sum_{\ell = 1}^{L_k} ||\bxi_{k,\ell}||^2$ and
\begin{align}\label{SE}
   	&\tilde{R}_\rsum  \notag \\
   	&= \log_2 \det\left( \bR_\bz +\sum_{k=1}^{K} \sum_{\ell = 1}^{L_k} \bH_{k}\bXi_{k,\ell}\bH_{k}^H\right) - \log_2\det\left( \bR_\bz \right).
\end{align}

\begin{figure*}[htbp]
	\centering 
	\vspace{-0.35cm}
	\setlength{\abovecaptionskip}{0.5cm}
	\includegraphics[width=1\linewidth]{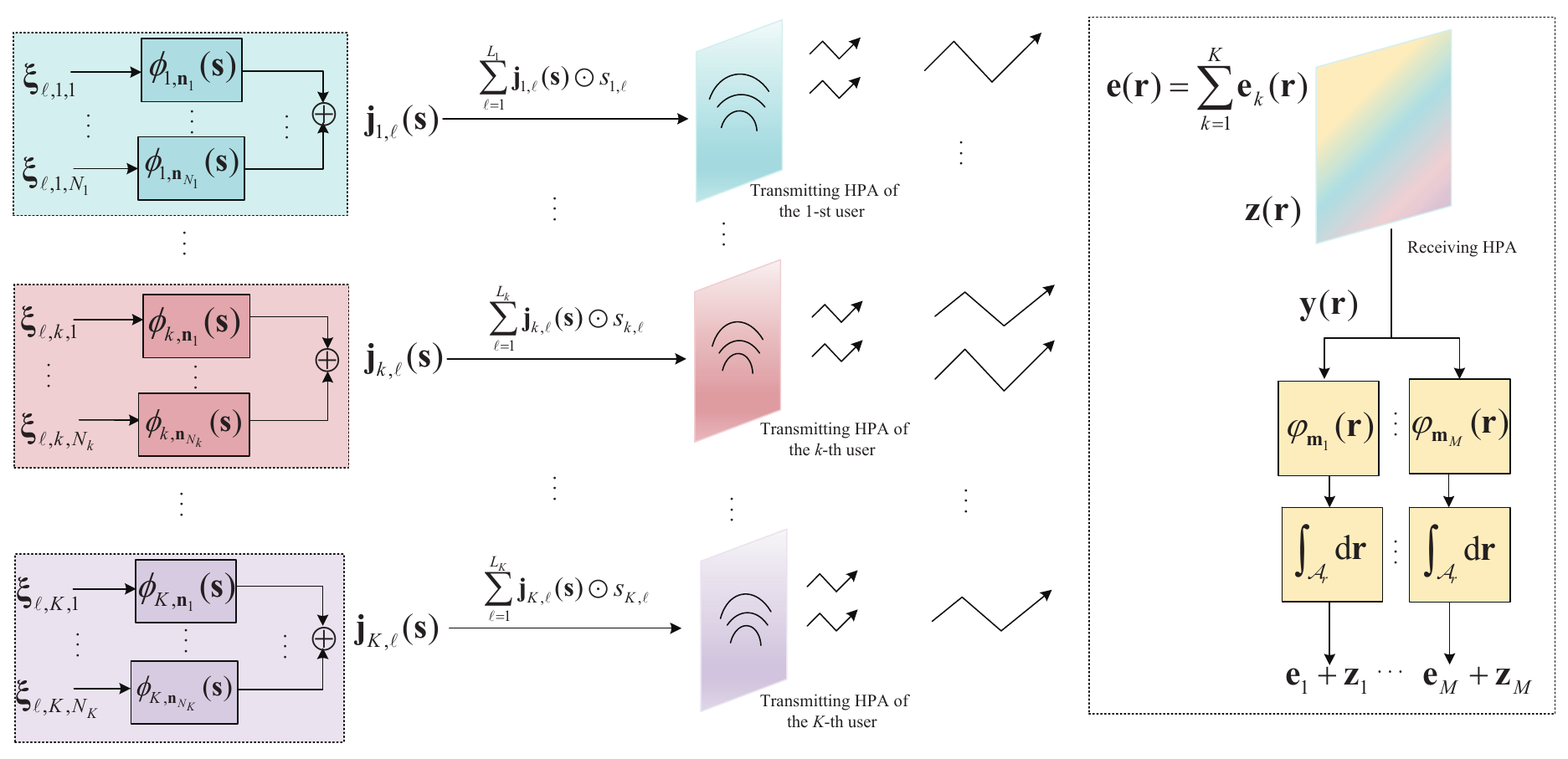}
	\caption{Transmission and receiving process of HPAs-assisted multi-user uplink transmission system.}
	\label{Modulation of J}
\end{figure*}

The signal processing procedures are summarized in \figref{Modulation of J}. 
By projecting the input signal $\bj_{k,\ell}(\bs)$ into a finite-dimensional space, the optimization variables become the finite-dimension vectors $\{\bxi_{k,\ell}\}_{\forall k,\ell}$. It is noteworthy that problem $\cP_{\tilde{A}}$ is distinct from the traditional problem of maximizing the SE in MIMO system for the following reasons.
Firstly, $\bH_{k}$ can be viewed as the sampling of the Green's function  in the wavenumber domain.
 Its $(j,i)$-th block, $\bH_{k,j,i} \in \bbC^{3\times 3}$,  represents the coupling matrix between the transmitting wave mode $i$ and the receiving wave mode $j$.  
Secondly, 
the dimensions of the optimization variables $\{\bxi_{k,\ell}\}_{\forall k,\ell}$ are related to the number of truncated Fourier space series items, i.e., $N_k$, which is decided by the dimension of the transmitting array size and the wavelength.   
In fact, for the $k$-th user, $\bxi_{k,\ell} \in\bbC^{3N_k\times 1}$ is the vector of projection coefficients whose dimension determines the maximum number of data streams that can be sent from the transmitting region, which means we should let $L_k \leq N_k$.


Since the term $\log_2 \det \{\bR_\bz\}$ does not affect the problem-solving,  we will omit it in the following optimization and the problem becomes
\begin{subequations}
\begin{align}
	\cP_{\hat{A}}: \quad \max_{\bXi_{k,\ell}}& \quad   \log_2 \det\left( \bR_\bz +\sum_{k=1}^{K} \sum_{\ell = 1}^{L_k}\bH_{k}\bXi_{k,\ell}\bH_{k}^H\right) \\
    \st& \quad  \sum_{\ell = 1}^{L_k}\trace{\bXi_{k,\ell}}  \leq P_{k,\max},  \quad \forall k\in \cK.
\end{align}
\end{subequations}
Let $\bXi_k = \rdiag{\bXi_{k,1}, \bXi_{k,2},\dots,\bXi_{k,L_k}}$ and $\hat{\bH}_k =  [\bH_k,\bH_k,\dots,\bH_k]$. Then, the problem can be re-formulated as 
\begin{subequations}
\begin{align}
		\cP_{\overline{A}}: \quad \max_{\bXi_{k}}& \quad   \log_2 \det\left( \bR_\bz +\sum_{k=1}^{K} \hat{\bH}_{k}\bXi_{k}\hat{\bH}_{k}^H\right) \\
	   \st& \quad \trace{\bXi_k}  \leq P_{k,\max},  \quad \forall k\in \cK.
\end{align}
\end{subequations}


At the optimal point, where each user's covariance matrix entails a water-filling of noise combined with interferences from all other users, it is reasonable to anticipate that an iterative algorithm can be employed to determine the optimal covariance matrices $\{\bXi_{k,\ell}\}_{\forall k, \ell}$ for the system SE.The specific procedures are as follows.

For the $k$-th user, denote the equivalent noise $\bB_k = \bR_\bz  + \sum_{k' = 1, k' \neq k}^{K} \hat{\bH}_{k'}\bXi_{k'}\hat{\bH}_{k'}^H $, which can be decomposed as 
\begin{align}\label{equivalent noise}
	\bB_k = \bU_k \bLambda_k \bU_k^H.
\end{align}
Then, the equivalent channel matrix $\tilde{\bH}_k = \bLambda_k^{-\frac{1}{2}}\bU_k^H\hat{\bH}_k$ and perform its singular value decomposition (SVD), 
\begin{align}\label{SVD on equivalent channel}
	\tilde{\bH}_k = \bF_k \bSigma_k \bT_k^H.
\end{align}

Thus, we optimize $\tilde{\bXi}_k \triangleq \bT_k ^H \bXi_k \bT_k$ to get the optimal solution of the problem, which is a diagonal matrix and  the $i$-th diagonal element is \cite{1262622}
\begin{align}\label{equivalent single-user water-filling algorithm}
	\tilde{q}_{k,i} = \left(\mu_k - \frac{1}{\sigma_{k,i}^2}\right)^+,
\end{align}
where  $\sigma_{k,i}$ is the $i$-th element of the diagonal matrix $\bSigma_k$ and $\mu_k$ is set to satisfy the power constraint $\trace{\tilde{\bXi}_k} \leq P_{k,\max}$. $(a)^+$ indicates that a larger value between $a$ and 0 is taken.
The algorithm pseudo code is summarized in \alref{alg: algorithm1}.

\begin{algorithm}
	\renewcommand{\algorithmicrequire}{\textbf{Input:}}
	\renewcommand{\algorithmicensure}{\textbf{Output:}}
	\caption{Iterative water-filling algorithm for HPAs-assisted uplink transmission system power allocation}
	\label{alg: algorithm1}
	\begin{algorithmic}[1]
		\REQUIRE Green functions $\bG(\br,\bs), \forall \br \in \cA_r  \text{ and }  \forall \bs \in \cA_{s,k} $; 
		Fourier space series $\phi_{\bn_{k,i}}(\bs), \forall i \text{ and }  \forall \bs \in \cA_{s,k} $; $\psi_{\bbm_j}(\br)$ $\forall j$ and $\forall \br \in \cA_{r}$; Threshold $\epsilon$.
		\STATE Initialization:$\left\{ \bXi_{k}^{(0)} \right\}_{\forall k}$ and set the iterative index $u = 0$.
		\REPEAT
		\STATE Calculate the system SE $R_\rsum^{(u)}$ based on \eqref{SE}.
		\FOR{$k = 1$ to $K$}
		\STATE Perform the eigenvalue decomposition on $\bB_k^{(u+1)}$ based on \eqref{equivalent noise}.
		\STATE Obtain the optimal $\tilde{\bXi}_k^{(u+1)}$ based on \eqref{equivalent single-user water-filling algorithm}.
		\STATE Obtain the corresponding $\bXi_k^{(u+1)} = \bT_k^{(u+1)} \tilde{\bXi}_k^{(u+1)} (\bT_k^H)^{(u+1)}$ by utilizing $\bT_k^{(u+1)}$ in \eqref{SVD on equivalent channel}.
		\ENDFOR
		\STATE Set $u = u+1$.
		\UNTIL{$|R_\rsum^{(u+1)} - R_\rsum^{(u)} |\leq \epsilon$}
		\ENSURE The maximal system SE $R_\rsum$.
	\end{algorithmic}  
\end{algorithm}

\subsection{Complexity and Convergence Analysis}
The main complexity of the algorithm lies in performing  SVD of the equivalent channel $\tilde{\mathbf{H}}_k$ after the eigenvalue decomposition of the equivalent noise $\mathbf{B}_k$. Both of these operations have a computational complexity of $\mathcal{O}(n^3)$, where $n \propto N_k$ is the dimension of $\mathbf{B}_k$ and $\tilde{\mathbf{H}}_k$, $\forall k \in \mathcal{K}$. Assuming that the algorithm requires a total of $I_{\mathrm{iter}}$ iterations, the overall complexity of the algorithm is $\cO(I_{\mathrm{iter}}Kn^3)$.

We would like to clarify the convergence of the algorithm as follows: at each step, the iterative water-filling algorithm finds the single-user water-filling covariance matrix for each user while regarding all other users' signals as additional noise. Since the single-user rate objective differs from the multi-user rate-sum objective by only a constant, the rate-sum objective is non-decreasing after each water-filling step. The rate-sum objective is bounded above, so the rate-sum converges to a limit \cite{1262622}. We admit that the input covariance \{$\bXi_{k}$\} obtained by the iterative water-filling algorithm may not unique since they depend on the initial value. However, they all obtain the optimal sum-rate value. 
The iterative water-filling algorithm demonstrates superior efficiency in comparison to general-purpose convex programming algorithms. It achieves convergence to the correct value with very few iterations, and its asymptotic convergence is exponentially rapid \cite{1262622}. We will provide the simulation results in Section \ref{sec:Simulation} to depict the convergence of the system SE versus the number of iterations for the proposed algorithm.

\section{Simulation Results}\label{sec:Simulation}
In this section, we will evaluate the proposed approach for our considered HPAs-assisted multi-user uplink transmission.
Throughout the simulations,  the system settings are generated in the following manner.  We set the number of the users $K = 4$. Unless explicitly specified, we maintain a constant EM wave frequency of 10 GHz and the number of data streams $L_k = N_k, \forall k \in \cK$.  The power densities  of both the hardware noise and the EM interference are $\frac{N_0}{2} = \sigma_{\remi}^2 = 5.6\times 10^{-6} \: \mathrm{V^2/m^2}$ \cite{sanguinettiWavenumberDivisionMultiplexingLineofSight2023}. 
Both the base station and the users are equipped with HPAs.  If not specified, we set the users on a semicircle centered on the base station with a radius length of  75 $\mathrm{m}$. 
It is worth noting that  the dimensions of the HPAs at the users vary between $1\lambda$ and $10\lambda$ in this simulation,  which are quite small in comparison to the propagation distance. 
Thus, we can safely assume that all users are equidistant from the base station.

The benchmark is set as follows. We assume that the transmitting antenna array of each user as well as the receiving array of the base station are planar arrays of discrete antenna units with the antenna spacing being $\lambda/2$ \cite{sanguinettiWavenumberDivisionMultiplexingLineofSight2023}. Besides, we assume that the current density function of each antenna unit is kept constant in both magnitude and phase, i.e., 
\begin{align}\label{discrete}
	\bj_{k,\ell} = \sum_{i = 1}^{T_{k}} \bbone_{\cA_{s,k}}(\bs \in \cA_{s,k}) \bbf_{k,\ell,i}, \quad k \in \cK,
\end{align}
where $\bbf_{k,\ell,i} \in \bbC^3$ is the precoding vector for the $\ell$-th symbol of the $i$-th antenna element of the $k$-th user. $T_k = T_{k,v} \times T_{k,h} $ is the total number of antenna units in the transmitting array for the $k$-th user and  we will use $R_{\mathrm{tot}} = R_{\mathrm{tot},x} \times R_{\mathrm{tot},y}$ to denote the number of antenna units in the receiving array. Since we set the spacing of each antenna unit  to be $\lambda/2$, $T_{k,v} = \lceil 2 L_{k,v}/\lambda \rceil$, $T_{k,h} =\lceil  2L_{k,h}/\lambda \rceil$, $R_{\mathrm{tot},x} = \lceil  2 R_x/\lambda \rceil $ and $R_{\mathrm{tot},y} =\lceil  2 R_y / \lambda \rceil$, with $L_{k,v}$ and $L_{k,h}$ being the length and the width of the $k$-th user's  transmitting HPA, respectively.
As a benchmark, we will optimize the precoding vectors $\left\{\bbf_{k,i}\right\}_{\forall k,\ell,i}$ to obtain the maximal system SE in this case.

\subsection{Comparison between the Proposed Scheme and the Benchmark Schemes}
\begin{figure}[htbp]
	\centering 
	\includegraphics[width=0.9\linewidth]{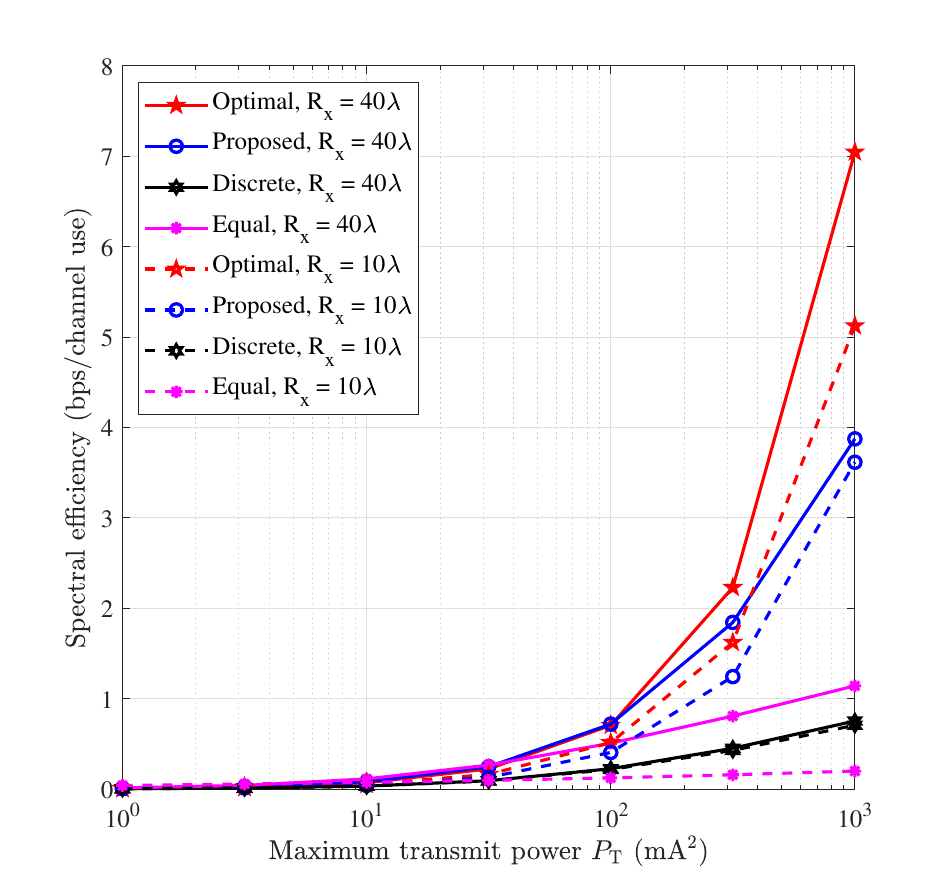}
	\caption{Comparison of the SE performance versus $P_{T}$ with different  optimization schemes.}
	\label{proposed_benchmark}
\end{figure}
\begin{figure}[h]
	\begin{center}
		\includegraphics[width=0.9\linewidth]{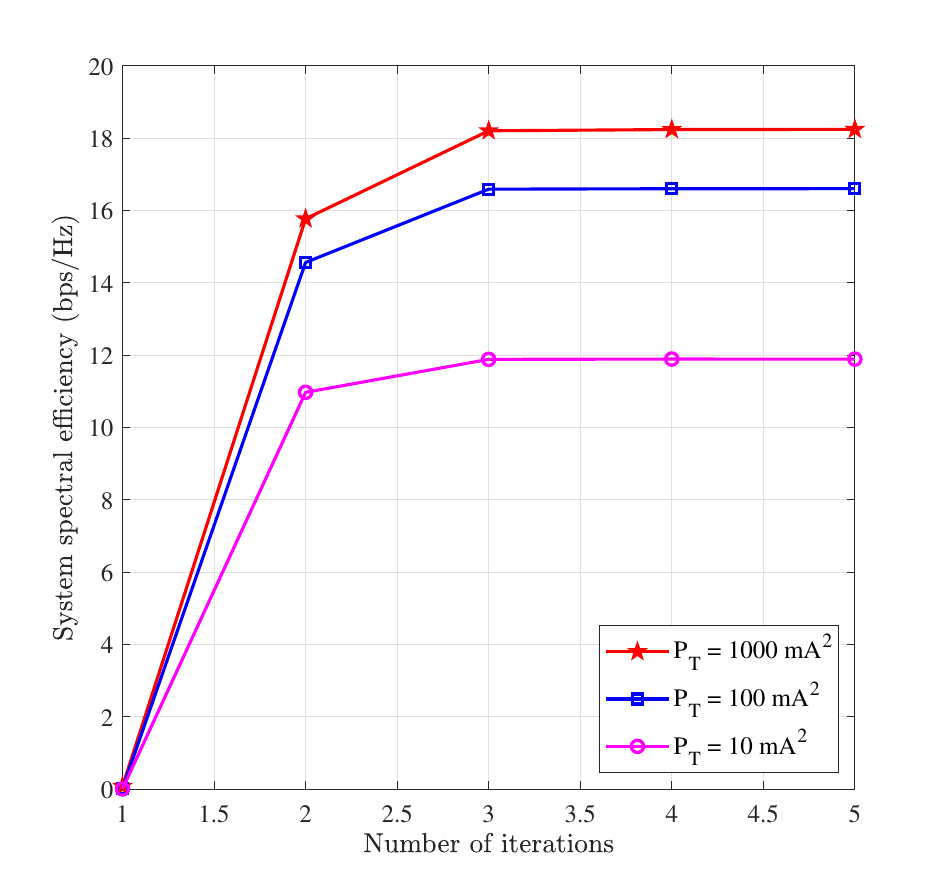}
		\caption{Average convergence performance versus the number of iterations.}
		\label{convergence}
	\end{center}
\end{figure}

\figref{proposed_benchmark} compares the system SE of the proposed algorithm and the benchmark schemes for different transmitting powers.
We fix $S_{k,x} \times S_{k,y}, \forall k \in \cK$, to $2\lambda\times2\lambda$. 
 ``Discrete''  means the benchmark scheme mentioned in \eqref{discrete} while ``Equal'' represents the scheme that distributes the transmitting power equally across different communication modes.  ``Optimal'' is the scheme that uses the optimal decomposition mentioned in \cite{millerCommunicatingWavesVolumes2000}.
 From the simulation results 
we can observe that  the gains obtained by the proposed scheme are much greater compared to the ``Discrete'' and ``Equal'' schemes, especially when the receiving size is larger such as $40\lambda \times 40\lambda$. For a discrete antenna array with $\lambda/2 $ spacing between antenna units, we observe that increasing  receiving array size will not improve the system SE. 
This is due to the fact that for traditional discrete array antenna systems, the DoF corresponds to the rank of the channel matrix, which is always no larger than the minimum between the numbers of transmitting antennas, $N_t$, and receiving antennas, $N_r$ \cite{tse2005fundamentals}. Therefore, changing the receiving array size does not provide any gain in the system SE for the discrete MIMO system since the transmitting array size at the user end is usually smaller than that at the receiving end. However, for systems assisted by HPAs, the effective DoF  is not $\min\{N_t,N_r\}$ even in far-field scenario \cite[Eq. (24)]{dardariCommunicatingLargeIntelligent2020}. Regardless of whether the receiving and transmitting arrays are parallel or orthogonal to each other, we can see that the system DoF is related to the size of the receiving area. So increasing the size of the receiving area can increase the system DoF gain, which in turn improves the system SE. It can be observed that the gap between the optimal scheme and the proposed scheme is quite small when the transmitting power is limited. While there remains a performance gap between the proposed solution and the optimal one,  the gap is acceptable especially considering that the proposed solution does not entail high decomposition complexity compared to the optimal solution.

\figref{convergence} shows the system SE obtained using the proposed algorithm versus the number of the iterations. It can be observed that for a wide range of the transmitting power, the system SE achieved by the proposed algorithm remains essentially unchanged after a few iterations. This indicates that the proposed algorithm converges rapidly.

\subsection{The Impact of the System Parameters}
\begin{figure}[htbp]
	\centering 
	\vspace{-0.35cm}
	\setlength{\abovecaptionskip}{0.5cm}
	\subfloat[Comparison of the SE performance versus $P_{T}$ with different wave frequencies and propagation distances.]{
		\includegraphics[width=0.9\linewidth]{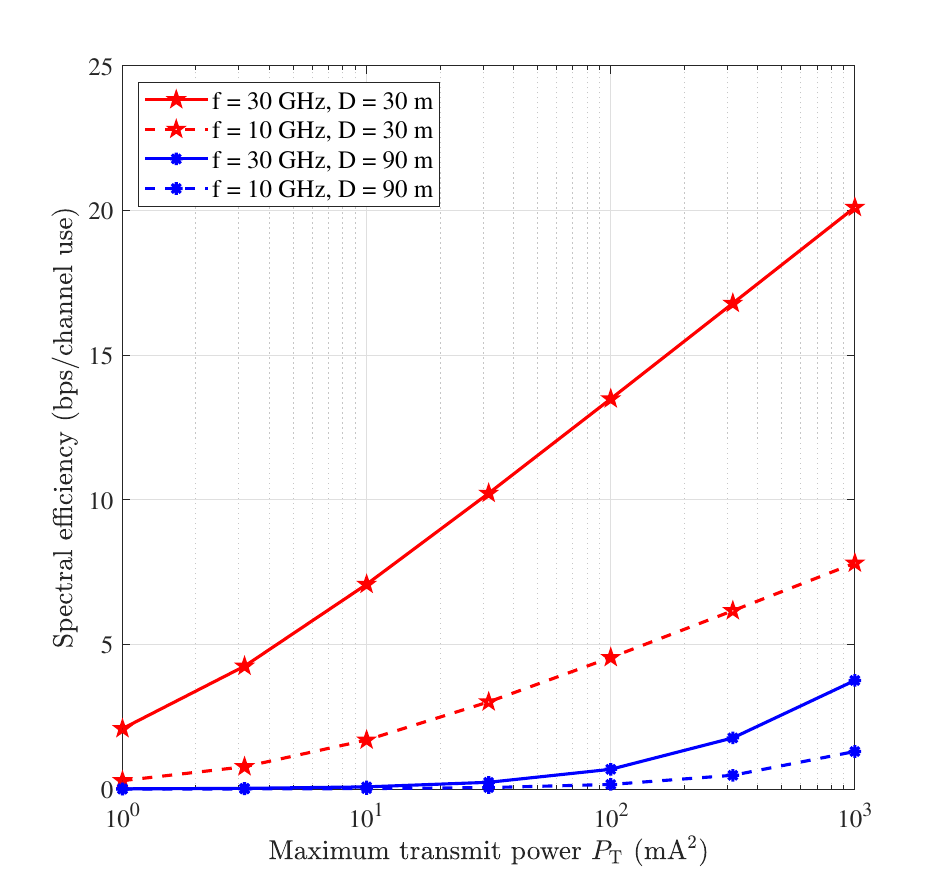}
		\label{distance_frequency_plot}
	}\quad
	\subfloat[Comparison of the SE performance versus $P_{T}$ with different array sizes.]{ 
		\includegraphics[width=0.9\linewidth]{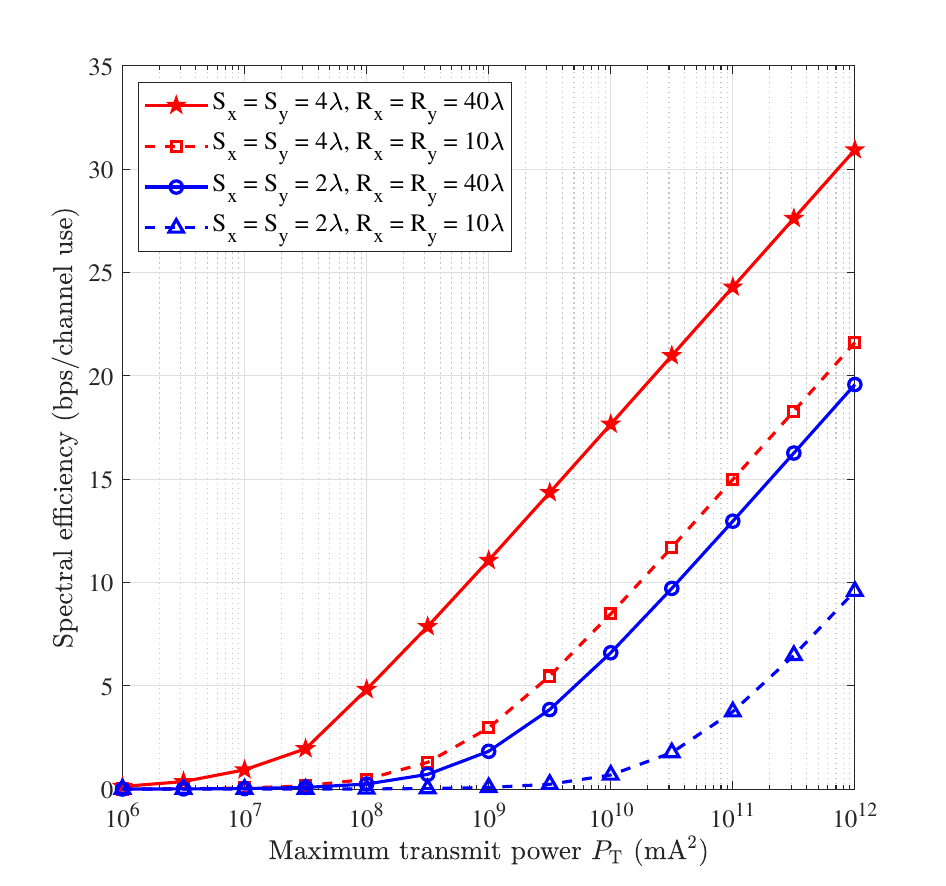}
		\label{array size_plot}
	}
	\caption{The impact of different system parameters on the system SE.}
	\label{system_parameters}
\end{figure}
We will explore the impact of system parameters such as array size, propagation distance and wave frequency on the system SE. It is worth noting that for continuous arrays, the ``array size'' refers to the size of the transmitting (or receiving) area.

First, keeping the same transmitting and receiving array size, we explore the effects of wave frequency and propagation distance on the system SE. From Fig.  \subref*{distance_frequency_plot} we can see that increasing the  frequency does help to improve the SE of the system. It is due to the fact that under the same array size and the propagation distance, the higher the wave frequency is, the greater the spatial freedom of the array is, since the effective DoF is inversely proportional to the wavelength $\lambda$ \cite{dardariCommunicatingLargeIntelligent2020}. Thus, the number of data streams that can be propagated independently is larger, which results in a higher SE of the system.  In addition, the variation of the propagation distance also affects the SE of the system, because the smaller the distance, the stronger the coupling between the received and transmitted signals, i.e., the less energy is dissipated at other locations.

Next, we investigate the effect of the transmitting array size and the receiving array size on the system SE. 
Since we are considering the effect of the transmitting array size on the system SE, we make a slight abuse of notation here, i.e., we let the $S_x$ and $S_y$ labeled in Fig. \subref*{array size_plot} denote the length and width of the user's HPA, respectively. In this way, since the user's position remains unchanged and so does its orthogonal transformed matrix $\bU_k$, by increasing (or decreasing) $S_x$ and $S_y$, the values of its projections along the $x$ and $y$ axes in the coordinate system parallel to the receiving plane are increased (or decreased) accordingly. Note that we set the transmitting array size to be the same for each user and change their sizes at the same time.
From Fig. \subref*{array size_plot}, it can be seen that an increase in both the transmitting array size and the receiving array size is beneficial to improving  the system SE. However, in practical applications, too large an array size will consume too many spatial resources, so a compromise is always needed.  The former will increase the system  spatial DoF, while the latter will increase the coupling strength. 
Although the increase in coupling strength often has an upper limit for improving the system SE, in the practical scenario where the receiving array size is limited, both contribute to enhancing the system SE.

\subsection{The Impact of the Colored Noise}

\begin{figure}[htbp]
	\centering
	\includegraphics[width=0.9\linewidth]{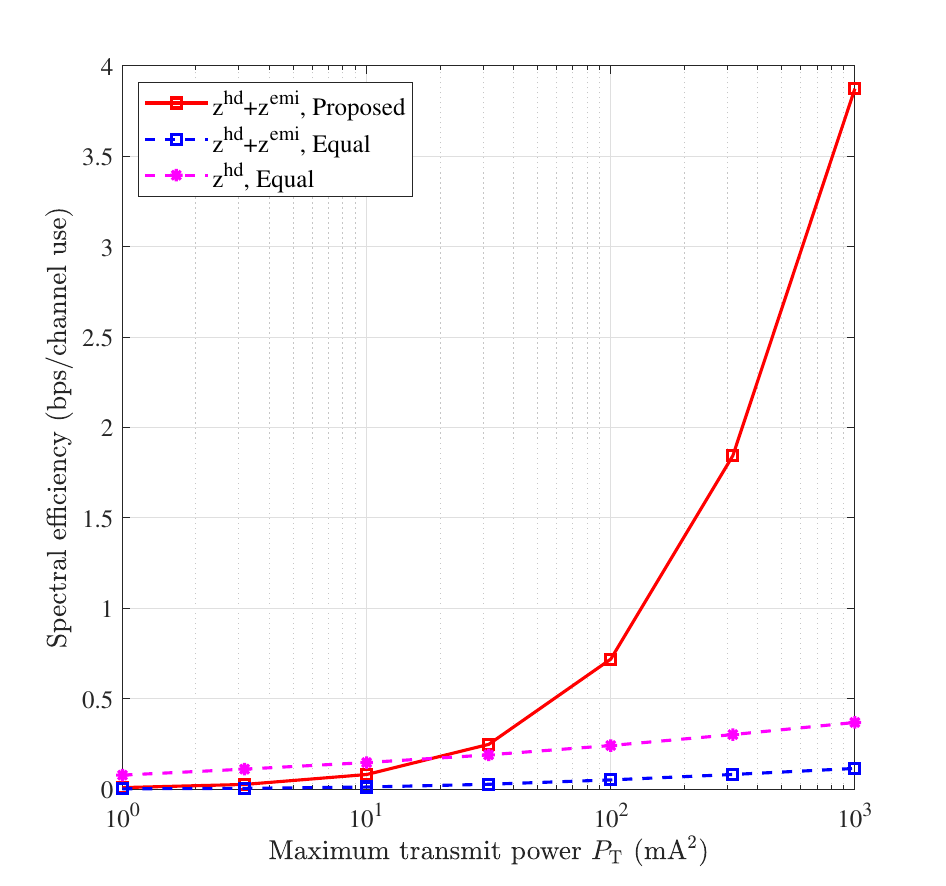}
	\caption{Comparison of the SE performance versus $P_{T}$ with different noise types.}
	\label{colored_noise_impact}
\end{figure}

 \figref{colored_noise_impact} demonstrates the advantage of the proposed scheme for the system SE improvement while considering colored noise. When doing power averaging along each communication mode, the SE in the case of only hardware noise (which is labeled by ``$\bz^{\mathrm{hw}}$'')  is  bigger than that of considering both hardware noise and EM interference (which is labeled by ``$\bz^{\mathrm{hw}}+\bz^{\mathrm{emi}}$'') . However, after the optimized design, the system SE  of colored noise case will be improved significantly.  While incorporating EM interference into consideration is more in line with physical reality,  the use of the proposed algorithm will significantly enhance the system SE in this case.

\section{Conclusion}\label{sec:conclusion}
In this paper, we have studied the uplink transmission for the multi-user HPA-assisted system. 
First, considering both the users and the base stations equipped with HPAs, we have developed a multi-user uplink transmission  model  including the EM channel model and the colored noise model, 
which considers hardware noise and EM interference.
Subsequently,  given that the system model in continuous space may entail significant computational complexity for analyzing and optimizing system performance, we have approximated the system model to discrete space using Fourier series expansion. 
We have shown that the EM channel model can be viewed as a sampling of the dyadic Green's function.
Based on the finite-dimension  models, we have formulated the problem of maximizing the SE for the  HPAs-assisted multi-user uplink transmission system,
for which an iterative water-filling algorithm has been proposed for the optimal  power allocation matrices. Finally, we have compared the advantages of the proposed scheme against  two benchmark schemes and clarified the impact of the colored noise and the system parameters on the SE through simulation results.

%
%
%
%
\bibliographystyle{IEEEtran}
\bibliography{Refabrv_20180802.bib, ref_HMIMO.bib}
\end{document}